# ATOM-WALL INTERACTION

**Daniel BLOCH and Martial DUCLOY**

*Laboratoire de Physique des Lasers, UMR 7538 du CNRS, Université Paris13*

*99 Av JB Clément , F- 93430 Villetaneuse, FRANCE*

**Abstract**

This chapter deals with atom-wall interaction occurring in the "long-range" regime (typical distances: 1-1000 nm), when the electromagnetic fluctuations of an isolated atom are modified by the vicinity with a surface. Various regimes of interaction are discussed in an Introductory part, from Cavity Quantum ElectroDynamics modifications of the spontaneous emission, to Casimir effect, with emphasis on the atom-surface van der Waals interaction, characterized as a near-field interaction governed by a $z^{-3}$ dependence. The major part of the Chapter focuses on the experimental measurements of this van der Waals interaction, reviewing various recent techniques, and insists upon optical techniques, and notably selective reflection spectroscopy which is particularly well-suited when excited atoms are considered. A review of various experiments illustrates the specific effects associated with a resonant coupling between the atomic excitation and surface modes, from van der Waals repulsion to surface-induced resonant transfer, and with anisotropy effects, including metastability transfer induced by a quadrupole contribution in the interaction. The effects of a thermal excitation of the surface -with a possible remote energy transfer to an atom-, and of interaction with nanobodies -which are intrinsically non planar- are notably discussed among the prospects.

## 1. <u>Introduction</u>

## 2. <u>Long range atom-surface interaction: principles and near-field limit</u>

**2.1** *Ground state atom and perfect reflector: from London-van der Waals interaction to the retarded Casimir-Polder limit*

**2.2** *Excited atom in front of a reflector : Radiative and $z^{-3}$ near field behaviors*

**2.3** *vW surface shift and virtual transitions*

**2.4** *Interaction with a dielectric medium*

**2.5** *Near-field modification of the lifetime of an excited atom in front of a real surface*

**2.6** *Interaction of an atom with an anisotropic medium*

## 3. <u>Experimental approaches for the probing of atom-surface interaction</u>

**3.1** *Observation of the vW interaction through mechanical effects*

  **3.1.1** <u>Thermal beam</u>

  **3.1.2** <u>Cold atoms</u>

**3.2** *Probing the vicinity of a surface with Selective Reflection spectroscopy*

  **3.2.1** <u>Basic principle</u>

  **3.2.2** <u>Atomic response and atomic motion</u>

    *3.2.2.1 The resonant atomic response*

    *3.2.2.2 The SR lineshape and the sub-Doppler logarithmic singularity*

    *3.2.2.3 The narrowing of SR lineshape with FM modulation*

  **3.2.3** <u>SR lineshapes in the presence of an atom-surface interaction potential</u>

    *3.2.3.1 The general case*

    *3.2.3.2 The specific case of the $z^{-3}$ vW interaction*

  **3.2.4** <u>Beyond several simplifying approximations</u>

    *3.2.4.1 Absorption*

    *3.2.4.2 Symmetry of the velocity distribution*

    *3.2.4.3 De-excitation at the arrival onto the surface*

    *3.2.4.4 Finite Doppler width*

    *3.2.4.5 Normal and oblique incidence*

    *3.2.4.6 Atomic trajectories: Spectroscopy vs. mechanical effects*

**3.3** *Nonlinear Selective Reflection*

  **3.3.1** <u>Saturation with a single irradiating beam</u>

**6.** **<u>Conclusion</u>**

1.    **Introduction**

The van der Waals (vW) attraction between two bodies is an ubiquitous phenomenon in nature, notably important to explain cohesion properties of materials. It originates in the correlation between the unavoidable electromagnetic (e.m.) field fluctuations (of quantum or thermal origin) of the two facing bodies. In spite of its extreme importance, direct physical measurements of this fundamental attraction are actually scarce. These effects are intimately connected to the Casimir attraction (Casimir 1948), a paradigm essential  for Quantum Electrodynamics (QED) (for a review see *e.g.* Bordag *et al.*, 2001), and of fundamental importance in many problems, including the determination of the cosmological constant. The Casimir force describes the attraction between two reflecting surfaces due to vacuum fluctuations and it includes the retardation effects (light propagation). The problem of two media separated by vacuum, can be viewed as a generalized vW attraction. Very recently, precise measurements of the Casimir attraction have appeared (for a review see *e.g.* Lambrecht and Reynaud, 2003), that stimulate new developments in the theory, notably justifying studies at various distance ranges, or consideration of material dispersion effects and thermal effects.

An elementary situation for this interaction between two bodies is the restriction to the interaction between a surface and a single atom. The theoretical link between the two problems goes through the approximation of an infinitely dilute medium (see Barash and Ginzburg, 1989). One experimental advantage is the variety of investigation methods that becomes available : among others, molecular beam technology, high resolution spectroscopy, laser cooled atoms, have been found appropriate for these studies. At the other end, the atom-surface problem can be seen as an extrapolation, through an integration over the collection of atoms that is the essence of the dense media, of the atom-atom long-range interaction, known as the vW long-range tail of an interatomic potential (London, 1930 ; Lennard-Jones, 1932).

More generally, the long-range atom-surface interaction is a topic of interest in its own right. "Long-range" here means that the atom is at such a distance from the surface -usually ≥ 1 nm-, that the atom does not feel the details of the atomic structure at the surface : the physical surface can be simply approximated as a planar *wall*, and a 2D symmetry translation is naturally introduced in the problem. This atom-wall interaction is an unvoidable feature in Cavity QED, a realm of Physics that has demonstrated the possibility of reversible exchanges of excitation between the cavity modes and the atom (see *e.g.* Haroche, 1992), or that permits such spectacular effects as the enhancement/inhibition of the atomic spontaneous emission with respect to the atom-surface distance (Kleppner, 1981; Heinzen *et al.*, 1987). In addition, with emerging nanotechnologies and their possibility of atom-by-atom implantation on a substrate, the precise knowledge of the interaction governing an atom in its approach towards a surface has become an essential concern. Similarly, for the understanding of desorption processes, it should be essential to know how a departing atom evolves from the attractive trapping region to the free-space.

The knowledge of the atomic structure, at least in the free-space, is usually extremely precise. This permits a detailed description of the atom-surface interaction provided the dense medium itself is described not in an ideal manner, but realistically (dispersive dielectric medium, real metal, ...). The present work mostly concentrates on the interaction of an *excited* atom with a surface. This situation, although more specific, is of an obvious interest for various applications, notably nanochemistry, and is susceptible to offer a variety of behaviors. As will be shown, it provides also a deepened insight into the effect of a possible excitation of the surface itself.

In addition to the energy-shift induced by the attractive potential exerted by the wall, one predicts for excited atoms a surface-induced modification of the atomic lifetimes. Drexhage and coworkers extensively studied this problem as early as the 70's, in the experimental situation of atomic species ($Eu^{3+}$ notably) actually embedded in organic layers (Drexhage, 1974). An overall agreement was demonstrated between the experimental results and the theoretical predictions, that were actually based upon a model of isolated atom evolving freely at some distance from a surface. The extension to atom-surface interaction naturally permits to explore more singular and quantized systems than those available with embedded emitting species.

Among these general topics, we have been involved from the early 90's in the problem of the near-field interaction between an excited atom and a surface, with experiments most often relying on dedicated optical techniques. The theoretical developments have encompassed the effects of a dielectric resonance, anisotropy in the interaction, and shape factor for interaction with microbodies and nanobodies. This review first summarizes the essential relevant results of Cavity QED, notably in its connection with the non-retarded vW interaction, and discusses the various subtleties that often make the electrostatic approach oversimplified. The following section (section 3) reviews the main experimental methods. The emphasis is on the optical methods, mostly relying on the monitoring of the reflected light at an interface, and on the transmission through a very thin vapor cell. This description offers the basis to the analysis, in section 4, of some of our essential experimental results : the spectroscopic approach allows one to observe the atomic behavior at a typical distance to the surface in the ~100 nm range. Before the final conclusion, the section 5 deals with the most recent developments (both experimental and theoretical) that are in particular oriented towards the interaction of atomic systems with on the one hand, micro-and nano- structures, and on the other hand with thermal excitation of the surface.

## 2.    Long range atom-surface interaction: principles and near-field limit

The major principles that govern Cavity QED and interaction of an atom with a reflector, notably a perfect reflector, have been extensively studied in numerous works, including *e.g.* the review by Hinds (1994). The aim of this section is first to recall major results for the nonfamiliar reader, and to emphasize theoretical results that are the more specific to the experimental works reviewed in this paper. Hence, we mostly focus here on the near-field limit of the surface interaction, corresponding to the vW description, and then discuss with more details the specific case of an interaction with a real reflecting surface, describing the reflector as a dielectric medium instead of an ideal metal.

### 2.1 *Ground state atom and perfect reflector: from London-van der Waals interaction to the retarded Casimir-Polder limit*

An elementary approach of the atom-surface interaction can be traced back to Lennard-Jones (1932). It relies on the idea of a London-van der Waals dipole-dipole interaction (London, 1930) between an electric dipole $\vec{d}$, and its electrostatic image induced in a reflector (see Fig.1). The interaction potential is hence given by:

$$V = -\frac{1}{4\pi\varepsilon_0} \frac{(d^2 + d_z^2)}{16 z^3} \qquad (1)$$

In Eq. (1), z is the dipole-to-surface distance, and ($d_x$ , $d_y$ , $d_z$) the components of the vectorial dipole $\vec{d}$. Although an atom has in general a null permanent electric dipole (dipole quantum operator **D**, with <**D**(t)> = 0) , its quantum (quadratic) fluctuations cannot be neglected (<**D²**(t)> ≠ 0). This permits to extend the electrostatic model, taking into account the instantaneous correlated fluctuations induced in the reflector. One finds an additional interaction potential $H_{vW}$ to be included in the atomic Hamiltonian:

$$H_{vW} = -\frac{1}{4\pi\varepsilon_0}\frac{(\mathbf{D}^2 + \mathbf{D}_z^2)}{16\,z^3} \tag{2}$$

Equation (2) shows that the energy shift of the ground state induced by the surface interaction is always negative, - *i.e.* attraction- and is governed by a factor <g|$\mathbf{D}^2+\mathbf{D}_z^2$|g> (|g> : the ground state wavefunction). Remarkably, this surface interaction potential is not isotropic; however, the anisotropy vanishes in the common situation of an atom featuring a spherical symmetry, as most of atom ground states.

It has been soon predicted by Casimir and Polder (1948), in a work parallel to the Casimir study (1948) of the electromagnetic interaction between two metal plates, that this $z^{-3}$ behavior is not compatible at long distance with the non instantaneous propagation of light. Taking into account the retardation effects, they show that the actual behavior evolves continuously from a $z^{-3}$ dependence (for z→ 0) to an asymptotic $z^{-4}$ (for z → ∞), with a final coefficient becoming governed by the (isotropic) electric atomic polarizability. Experimental evidences of this deviation from $z^{-3}$ behavior to the $z^{-4}$ behavior at long distance have appeared only in the recent years (Sukenik *et al.*, 1993; Landragin *et al.,* 1996; Shimizu, 2001; for a review, see Aspect and Dalibard, 2003).

An important point recognized by Casimir and Polder (1948) with these retardation effects is that the distance range in which the instantaneous (near-field) approximation -also called electrostatic interaction- is applicable corresponds to distances smaller than the (reduced) wavelength of the relevant quantum transitions (see Fig. 2). For a real atom, and in spite of the many transitions that can couple the ground state to excited states, there are numerous situations for which a two-level approximation, limited to the ground state and the resonant level, is valid: it yields a simple estimate of the range for which the short distance approximation (*i.e.* instantaneous image) applies.

## 2.2 *Excited atom in front of a reflector : Radiative and $z^{-3}$ near field behaviors*

To describe the surface effects on an atom in an excited state, one needs to consider the radiative properties of the atom associated to spontaneous emission. This comes in addition to the quantum dipole fluctuations, that are similar in essence to those for a ground state. This spontaneous emission can be described as radiated by an atomic dipole oscillating at the transition frequency -or as a sum of such radiations, if various decay channnels are opened-. In the vicinity of a reflector, the boundary conditions are responsible for a "self-reaction" correction term that imposes modifications to the radiative diagram in vacuum. Indeed, the oscillating atomic dipole (object) interferes with the oscillating dipole image induced in the reflector, whose oscillation is now phased-delayed by propagation effects: this results in both a radiative energy shift, and a modified decay rate of the spontaneous emission.

The vacuum wavelength of the electromagnetic oscillation provides a natural scaling factor to evaluate the distance range to the surface, sorting far-field and near-field effects (see Fig.2). In agreement with the well-known far-field expansion of the energy radiated by an oscillating dipole (Hinds and Sandoghdar, 1991), one finds for an excited state a slowly damped oscillating behaviour, evolving asymptotically as $z^{-1} cos(kz)$ (k: the wavenumber associated to the considered radiative process), at odds with the Casimir-Polder limit applicable to a ground state. This oscillatory radiative shift, although it remains tiny relatively to the inverse of the spontaneous emission decay time, has been notably observed with an atomic beam passing through the node or antinodes of an orthogonal Fabry-Perot (Heinzen *et al.*, 1987). Conversely, in the near-field limit, the radiative shift is dominated by a $z^{-3} cos(kz)$ term. In the electrostatic limit ($kz \ll 1$), on which we will mostly concentrate on, this contribution turns out to be nothing else than the $z^{-3}$ vW shift mentioned above.

### 2.3 *vW surface shift and virtual transitions*

Before proceeding to the quantitative evaluation of the vW surface interaction, it may be of interest to note that a pure classical atom description, with an orbiting electron, already provides some modeling of the dipole atomic fluctuations. The more the orbiting electron is excited, the stronger are the fluctuations. As a result, optical spectroscopy can in principle be used to evidence an atom-surface interaction, with a $z^{-3}$ red-shift in the near-field electrostatic approximation. This has provided the basis for numerous experiments, such as evoked in the following sections. Paradoxically, a pure quantum two-level model contradicts this prediction (Hinds and Sandoghdar, 1991), as, in this limited frame, the average dipole fluctuations are equal for each level, and the spectral line is not shifted. This remark may help to understand the fundamental importance of considering virtual transitions in the evaluation of the strength of the $z^{-3}$ vW interaction. With the expansion $\mathbf{1} = \sum_j |j\rangle\langle j|$, (with $\mathbf{1}$ the identity operator), one gets:

$$\langle i|\mathbf{D}^2|i\rangle = \sum_j \langle i|\mathbf{D}|j\rangle \ \langle j|\mathbf{D}|i\rangle = \sum_j \left|\langle i|\mathbf{D}|j\rangle\right|^2 \tag{3}$$

where appears in Eq. (3) a sum over all the virtual transitions (a similar development holds for the non scalar contribution in $\mathbf{D}_z^2$).

Remarkably, parity considerations impose that when comparing $\langle i|\mathbf{D}^2|i\rangle$ and $\langle j|\mathbf{D}^2|j\rangle$ (with $|i\rangle$ and $|j\rangle$ connected by an allowed dipole transition), the virtual couplings relevant for each of the two levels $|i\rangle$ and $|j\rangle$ define two different sets of atomic levels : this confirms that a pure two-level model is essentially unable to predict the strength of the vW interaction. This strength can be conveniently described by a $C_3$ coefficient, related to the vW Hamiltonian $\mathbf{H}_{vW}$ by:

$$C_3(|i\rangle) = -\frac{z^3 < i|\mathbf{H}_{vW}|i>}{h} \tag{4}$$

( with $h$ the Planck constant), so that :

$$C_3(|i\rangle) = \frac{1}{4\pi\varepsilon_0} \frac{1}{16\,h} \left[ \sum_j \left|<i|\mathbf{D}|j>\right|^2 + \sum_j \left|<i|\mathbf{D}_z|j>\right|^2 \right] \tag{5a}$$

or, assuming an isotropic interaction (*e.g.* spherical symmetry for the considered $|i\rangle$level),

$$C_3(|i\rangle) = \frac{1}{48\pi\varepsilon_0 h} \sum_j \left|<i|\mathbf{D}|j>\right|^2 \tag{5b}$$

In general, the prediction for the $C_3$ value characterizing the vW interaction can be rather accurate, the atom being in the ground state or in an excited state atom. This is because Atomic Physics has developed numerous and refined tools for the evaluation of atomic wavefunctions. However, a few remarks should be pointed out:

(i) Oppositely to the spontaneous emission, which is negligible in the far infrared (IR) range, and dominant in the UV part of the spectrum, the $C_3$ value is often dominated by the contribution of dipole couplings associated to virtual (far) IR transitions (see *e.g.* Fig. 3 and Table 1). This is due to the $\lambda^3$ dependence appearing in the dipole coupling for a given oscillator strength.

(ii)   In most cases, the $C_3$ value grows with the atomic excitation level: this can be seen either from the increasing dipole fluctuations when the atomic excitation increases (comparable to a stronger atomic polarizability), either from the more tightened atomic structure, with more numerous IR couplings, typical of levels close to ionization.

(iii) The vW Hamiltonian $\mathbf{H}_{vW}$ is nonscalar in its essence, with a quadrupole term $\mathbf{D}_z^2$ susceptible to modify atomic symmetry. Such a modification becomes quite important when the vW interaction compares with the energy difference between levels, notably at very short distances (see section 5.1.3), or when some sublevel degeneracy is removed (*e.g.* in the presence of a magnetic field, with Zeeman degeneracy removed), or for a polarized atomic system. However, for a statistical set of sublevels, sum rules enable  the vW interaction to be simply estimated with the knowledge of the dipole fluctuations associated to the *radial* part of the wavefunction, and with simple *angular* rules (Chevrollier *et al.*, 1992). Note that in the principle, the various hyperfine components originating from the same level do not necessarily undergo an identical vW interaction potential, owing to their different angular momentum. However, for atomic states with low values of the angular orbital momentum (*e.g.* S or P levels, that can be coupled only to S, P, D levels) the anisotropy effect remains relatively small (see also section 4.5). More generally, as long as the magnetic component remains degenerate, the averaged vW shift is governed by the scalar vW contribution (that can be identified to the r.h.s. of Eq. (5b)), while the quadrupole contribution, that varies with the magnetic component, is essentially susceptible to induce an hyperfine-dependent broadening (Papageorgiou, 1994).

(iv) In the summing implied by Eq. (5),  the transitions to high-lying state and to auto-ionizing levels are very numerous, and their influence, in spite of their unfavorable short wavelength, can be important, notably for a ground state. Derevianko *et al.* (1999) have estimated that for a Cs atom in its ground state, the transitions involving the external electron contribute to only ~60 % of the vW interaction, in spite of the alkali nature of this atom. Indeed, a large part of the $C_3$ value originates in transitions involving excitation of the electronic core. However, the near-field approximation is valid only in a very limited range for these contributions, located in the vacuum ultra-violet (VUV) range. Above the range of core transition wavelengths, the near-field approximation (vW interaction) still stands, provided that the short wavelength couplings are neglected. Also, the core contribution does not vary much from one energy level to the other, and its contribution to spectral lines largely cancels.

## 2.4 *Interaction with a dielectric medium*

With respect to the large range of wavelengths involved in the vW coupling, no material can be expected to behave as an ideal reflector on the whole spectrum of interest. Rather, any real surface, including those made of noble metals, exhibits dispersive features, that can be dealt with when considering the general problem of an atom interacting (in the near-field approximation) with a dielectric medium.

In the frame of a pure electrostatic model, the induced image in a dielectric medium is simply reduced by an image coefficient r ($0 < r < 1$):

$$r = \frac{\varepsilon - 1}{\varepsilon + 1} \qquad (6)$$

with $\varepsilon$ the (real) dielectric permittivity of the medium. Actually, the static value of the permittivity, appearing in Eq.(6), is not relevant to deal with the real properties of the reflector. Indeed, in our problem of atom-surface vW interaction, the temporal dipole fluctuations have been introduced, and developed over a set of virtual transitions.

Following Lifchitz work (1956) for the interaction between two dielectric solid bodies, the general response for a ground state atom interacting with a surface has been approached in the 60's (Mavroyannis, 1963; Mac Lachlan, 1963 a, 1963 b). For each virtual transition $|i>$ $\rightarrow$ $|j>$ contributing to the vW shift, one must introduce in Eq.(5) an image factor $r(\omega_{ij})$ (with $\omega_{ij}$ : the transition frequency, counted $> 0$ for an $|i> \rightarrow |j>$ absorption):

$$C_3(|i>) = \frac{1}{48\pi\varepsilon_0 h} \sum_j r(\omega_{ij}) \left| < i|\mathbf{D}|j > \right|^2 \qquad (7a)$$

with :

$$r(\omega_{ij}) = \frac{2}{\pi} \int_0^\infty \frac{\omega_{ij}}{\omega_{ij}^2 + u^2} \frac{\varepsilon(iu) - 1}{\varepsilon(iu) + 1} du \qquad (7b)$$

In Eq. (7), $\varepsilon(iu)$ is the analytical extension to the complex plane of the frequency-dependent dielectric permittivity $\varepsilon(\omega)$. Equation (7) shows that for a fluctuation, apparently sensitive to a transition frequency $\omega_{ij}$, the response efficiency is dependent upon the whole spectrum of the material. Also, it can be shown from causality reasons (*e.g.* from the Kramers-Krönig relation), that $r(\omega_{ij})$ decreases in a monotonic way with increasing $\omega_{ij}$. This implies, in agreement with the pure electrostatic model, that $0 < r(\omega_{ij}) < 1$. Moreover, far away from dielectric resonances, $\varepsilon(\omega_{ij})$ is real and slowly varying, and $r(\omega_{ij}) \approx [\varepsilon(\omega_{ij}) - 1]/[\varepsilon(\omega_{ij}) +1]$ ;

this justifies that when the vW interaction depends only on transitions falling into the transparency region of the material, the image coefficient factor is approximated by r = (n² - 1) / (n² + 1), with n the refractive index of the material in the transparency region.

In most of the real situations, the transparency window of a material is not large enough (relatively to the width of the absorption bands), so that the influence of the neighboring absorption bands cannot be totally neglected. This brings some corrections to (n²-1)/(n²+1). This has been illustrated in the course of our work (Failache *et al*, 2003): while for sapphire, a material transparent from 0.2μm to 5μm, the estimated value at λ = 0.87 μm is r ∼ 0.49, the simple (n²-1)/(n²+1) approximation yields r = 0.51 .

When an *excited* state interacts with a dispersive reflector, the vW interaction can still be described by a set of dielectric reflection coefficients $r(\omega_{ij})$, but there is no longer such a restriction as $0 < r(\omega_{ij}) < 1$. From the Wylie and Sipe (1984 , 1985) approach, one gets indeed (Fichet *et al.*, 1995 a):

$$r(\omega_{ij}) = \frac{2}{\pi} \int_0^\infty \frac{\omega_{ij}}{\omega_{ij}^2 + u^2} \frac{\varepsilon(iu) - 1}{\varepsilon(iu) + 1} du + 2\Re e \left[ \frac{\varepsilon(|\omega_{ij}|) - 1}{\varepsilon(|\omega_{ij}|) + 1} \right] \Theta(-\omega_{ij}) \tag{8}$$

with $\Theta$ the Heavyside function [$\Theta(\omega) = 1$ for $\omega > 0$, $\Theta(\omega) = 0$ for $\omega < 0$)] and $\Re e$ standing for "real part". While the first term in the r.h.s. of Eq.(8) behaves as the contribution in Eq. (7), with an averaging over the whole spectrum of the material, the second term in the r.h.s. of Eq.(8) appears only for a *virtual emission* (*i.e.* for $\omega_{ij} < 0$), and has no equivalent for a ground state atom. It is calculated at the precise virtual transition frequency, and it is susceptible to take any real value, moreover without any sign restriction. The physics of this additional term relies on the possible resonant coupling between the virtual atomic emission, and a virtual absorption in a surface mode. The surface mode resonances, such as illustrated in Fig.4 for typical windows, are derived from the bulk resonances - governed by $\varepsilon(\omega)$ - , through a shape factor [($\varepsilon$ - 1) / ($\varepsilon$+1)]. Note that the resonant surface response can possibly induce a change of the sign of the atom-surface interaction -leading to a vW *repulsion*, instead of the universal attraction-. This can be interpreted in the following way : instead of instantaneous image fluctuations, the resonant coupling enables giant but time-delayed dipole fluctuations in the dense medium, as induced by the atomic fluctuations in the range of the virtual transition frequency. Note that as will be discussed later on (section 5), the restriction (in Eq. (8)) to a virtual atomic *emission* is related to the standard assumption that the surface itself does not bear excitation (*i.e.* the temperature is assumed to be T = 0). This is legitimate as long as far IR transitions can be neglected at room temperature.

## 2.5 *Near-field modification of the lifetime of an excited atom in front of a real surface*

The modifications of the lifetime induced in the vicinity of a reflector are also specific to the excited atom. For an *ideal reflector*, there appears, along with the damped oscillatory behavior of the energy shift (see section 2.2), an enhancement/inhibition of the spontaneous emission close to surface, depending on the dipole orientation: in the limiting situation z = 0, a dipole parallel to the wall does not radiate, because the induced image oscillates with a phase strictly opposite to the one of the dipole source, while the emission rate doubles for a dipole with a normal orientation. Such a behavior, easily derived from a classical model, has also been rigorously justified in the frame of Quantum Mechanics (Courtois *et al.*, 1996).

In front of a "real" surface, the above interference effects are attenuated because the reflection is only partial, but an additional damping must be considered, related with the opening of extra-decay channels. Indeed, for a transparent dielectric medium, and as discussed by Lukosz and Kunz (1977 a), the excited atom can lose its excitation by the emission, in an evanescent mode, of a photon, that has the ability to freely propagate in the transparent dielectric medium (see Fig. 5). This effect, associated to an atomic emission in the forbidden region outside the fluorescence cone, always enhances the atomic relaxation rate, but by an amount that remains finite, even very close to the interface. This amount depends on the dipole orientation and on the dielectric medium index, but it remains relatively small (*e.g.* the enhancement does not exceed a factor ~3 for an index n = 3, and for the optimal dipole orientation). Its variations with the distance to the surface, originating in the standard features of evanescent wave, typically span over the relevant wavelength. The extension to a multilevel atom, with various decay channels, is straightforward with respect to the various wavelengths to be considered.

When the wall is not transparent at the considered emission wavelength, the absorption prevents those extra-decay channels related with an emission propagating in the forbidden cone. However, energy can be transfered from the excited atom to the surface modes (e.m. propagation guided along the surface). This near-field coupling to a surface mode appears to be the dissipative counterpart of the vW energy shift. It is governed by a surface response that diverges as $z^{-3}$ (Wylie and Sipe, 1985), and governed by a factor $\Im m\left[(\varepsilon(\omega_{ij})-1)/(\varepsilon(\omega_{ij})+1)\right]$, whose variations for typical windows are illustrated in Fig.4. This divergence establishes the possibility of dramatic changes in the relative transition probabilities for an excited atom at a small distance from a wall (Failache *et al.*, 2002). Also, the efficiency of this transfer depends on the line strength: as for the vW interaction, the long wavelength transitions located in the far IR, where the spontaneous emission is usually negligible, are usually dominant in this near-field process.

### 2.6 *Interaction of an atom with an anisotropic medium*

The "long-range" approximation, as discussed in section 1, implies that the atom interacts in a system offering a perfect plane symmetry, so that the interacting Hamiltonian is governed by a cylindrical symmetry. Actually, this symmetry can disappear if the medium itself exhibits some anisotropy, *e.g.* if the interacting surface has finite dimensions or a spherical shape (see sections 5.3 and 5.4), or as discussed here, in the case of a birefringent medium. Up to a recent work, performed in our group (Gorza *et al.*, 2001) and involving a detailed treatment of the birefringent medium for an arbitrary orientation of the medium axis, this problem had been seldom tackled when the (quantized) interacting particle is an atom. Several studies, that had involved an interaction with liquid crystals, notably those of Kihara and Hondo (1965) and Bruch and Watanabe (1979), yield however valuable hints for an atom in the ground state (see also the work of Šarlah and Žumer, 2001). As a general result, a major modification in the interaction appears in the interaction Hamiltonian, now described by such an expansion:

$$\mathbf{H}_{vW} = -\frac{1}{4\pi\varepsilon_0}\frac{\alpha D_x^2 + \beta D_y^2 + \gamma D_z^2}{16 z^3} \tag{9}$$

with usually $\alpha \neq \beta$. This introduces in the principle extra quadrupole effects in the interaction, with the additional breaking of some usual symmetry rules. In the principle, this break of symmetry within the vicinity of the surface may enable previously forbidden transitions, analogous to those observed by Boustimi *et al.* (2001 c) and Karam *et al.* (2002).

Even for an atom with a low angular momentum, when the effects are restricted to the scalar part of the interaction, the medium anisotropy can still induce some sensitive effects, through an orientation-dependent shift of the resonant properties. Indeed, the dielectric properties of the medium are governed by two different permittivity coefficients (respectively the permittivity $\varepsilon_{//}$ along the birefringence axis, and $\varepsilon_{\perp}$ orthogonal to this axis), whose resonance frequencies are not equal. Hence, the extrapolated surface resonances, estimated from the surface response $[(\varepsilon-1) / (\varepsilon+1)]$ are also different (Fichet *et al.*, 1995; Gorza *et al.*, 2001; Failache *et al.*, 2003). Within the scalar approach, an effective dielectric permittivity $\varepsilon_{eff}$ can be defined, that averages the permittivities $\varepsilon_{//}$ and $\varepsilon_{\perp}$, with a geometrical weighting dependent on the orientation of the birefringent axis relatively to the surface. As illustrated in Fig.6 , the atom-surface resonant coupling becomes tunable through an adequate choice of the *c*-axis orientation, and this may permit to tailor a giant and possibly repulsive vW interaction, or to impose a specific near-field quenching of the atomic excitation.

## 3.     Experimental approaches for the probing of atom-surface interaction

Among the various experimental methods that have been used for the probing of atom-surface interaction, one may distinguish between techniques relying on mechanical effects, and methods based upon optical spectroscopy. There is in the principle no intrinsic limit to the spatial resolution of mechanical methods, while the optical spectroscopy methods are commonly plagued by a limited spatial resolution (on the order of $\lambda/2\pi \sim 100$ nm), that makes them non convenient to explore these smaller distances to the surface  (down to $z \geq 1$ nm) for which the surface interaction is described as "long-range". On the other hand, mechanical methods are only well suited to long-lived levels (*i.e.* ground state, or Rydberg states), while high-resolution spectroscopy is very convenient to study short-lived excited states.

### 3.1 *Observation of the vW interaction through mechanical effects*

#### 3.1.1 *Thermal beam*

The first observation related with the vW atom-surface interaction can be traced back to the 60's, with the Columbia group (Raskin and Kusch, 1969) evidencing the deflection of a thermal beam of atomic Cs, at a grazing incidence with a metallic cylinder. It was followed by comparable experiments with other species, including molecules, or with dielectric surfaces (Shih *et al.*,1974; Shih, 1974). However, in these pioneering experiments, a large distribution of  impact parameters is involved, while only atoms flying very close to the surface undergo a notable deflection. These measurements (Shih and Parsegian, 1975), affected by  the uncertainty on the quality of the polishing of the deviating cylinder, did not permit a sophisticated inversion of the potential, such as discriminating between a $z^{-3}$ vW potential, a $z^{-4}$ dependence (asymptotic limit of the Casimir-Polder expansion), and an hypothetical $z^{-2}$ potential (*e.g.* for a charged surface).

A series of beautiful experiments was completed by the Yale group with a beam of long-lived Na (and Cs) atoms in Rydberg states (Anderson *et al.*, 1988; Sandoghdar *et al.*, 1992 and 1996): essentially, the transmission of the atomic beam is measured when flying between two parallel surfaces, whose relative distance is adjustable. The high polarizability of Rydberg levels make them very sensitive to the vW attraction. The transmission methods provide a specific enhanced spatial resolution in the sense that only Rydberg atoms flying in a very narrow central zone, where the two vW attractions compensate, are susceptible to escape from the attraction of the two plates, and to be transmitted and finally counted through an ionization process. This favorable transverse velocity selection also justifies that a classical trajectory model reveals sufficient for the analysis. Precision measurement of the $z^{-3}$ behavior of the atom-surface interaction was achieved (Sandoghdar *et al.*, 1992 and 1996) by measuring the spectral shift of the excitation resonance to the Rydberg levels, for a beam of Na atoms having entered in the ground state in the channeling zone. The quantitative comparison was performed for a plate separation ranging from 3 μm down to a 0.5 μm value, and for various (nS) Rydberg levels (n = 10 to n = 13). It was found to be in agreement with the theoretical predictions. These experiments were essentially conducted for interaction with metal-coated walls. Attempts to extend these experiments to dielectric surfaces (notably with *uncoated* silica blocks) most often led to irreproducible results, with respect to the strong sensitivity of Rydberg levels to stray electric fields. Note that with slight modifications of the set-up, the long-distance Casimir-Polder behavior (for a ground state) have been demonstrated (Sukenik *et al.*, 1993), as well as modifications of the spontaneous emission induced by thermal effects on the wall (Lai and Hinds, 1998).

It is only very recently that more numerous experiments have been implemented, based upon a mechanical signature of the vW interaction. The technological development of nanogratings have enabled the Göttingen group (Grisenti *et al.*, 1999) to observe the modified atomic diffraction of a rare gas beam -in its ground state and more recently in a metastable state (Brühl *et al.*, 2002)- in a transmission grating. The vW interaction reduces the effective opening of the slits and induces a modification of the diffraction pattern, conveniently observed as due to the high quality and reproducibility of the nanoslits involved in the grating (periodicity ~ 100 nm). Relatively to the experiments on Rydberg levels, the small size of the transmission region is compensated for by the weaker interaction coefficient for a ground state. Metastable beam transmission in slits and gratings has also been used to explore the anisotropic part of the vW interaction (see section 5.3)

### 3.1.2 *Cold atoms*

The recent development of cold atom technology has opened up a new field to the study of atom-surface interaction. It is notably motivated by the fact that surface-induced effects, like decoherence, can be responsible for major limitations in the trapping and manipulation of cold atoms in integrated atom optics: in particular, at short distances, the vW attraction may be strong enough to attract cold atoms towards the surface and to accelerate them in a final phase of (surface) thermalization. However, the use of cold atoms for the study of atom-surface interaction have been rather restricted, most probably because of the unfavorable duty cycle between the duration required to prepare a bullet of cold  atoms (on the order of seconds) and the short time actually spent close to the surface (at 1 μm from the surface, the velocity largely exceeds ~0.1m/s , even  for a cold sample prepared 1 mm above the surface).

Anyhow, the Orsay group (Landragin *et al.*, 1996) has succeeded in probing the long-range atom-surface interaction, observing the free fall of a sample of cold atoms, down to the surface. They have provided one of the very few evidences of the retardation effects affecting the interaction with the ground state. Essentially, the attractive long-range potential exerted onto a ground state is balanced with a controlled  additional repulsive potential as induced by a (blue-detuned) evanescent wave. The resulting bouncing of atomic waves -occuring at a distance estimated to be  47 nm- appears to obey to the Casmir-Polder expansion, rather than to the simple  vW attraction, valid only for $z \ll \lambda$ (see section 2.1). It is worth noting by the way that, in the fitting, the atomic interaction is theoretically evaluated only with the help of the transitions involving the external electron, while neglecting  transitions from the core (Derevianko *et al.*, 1999). This is most probably justified (see section 2.3) by the distance of observation, that largely exceeds the wavelength (in the VUV range) of transitions from the core, hence inducing a strong attenuation through the retardation effects; in addition, the reflectivity of the dielectric surface is most probably very weak for the VUV range.

Further possibilities to explore a closer range of distance with cold atoms seems now opened with  the recent experimental demonstration of quantum reflection, performed by Shimizu (2001) with a beam of ultra-cold Ne* atoms flying nearly parallel to the wall.

### 3.2 *Probing the vicinity of a surface with Selective Reflection spectroscopy*

#### 3.2.1 *Basic principle*

The basic principle of the Selective Reflection (SR ) spectroscopy was devised nearly a century ago (Wood, 1909). It consists in the monitoring of a resonant change of the reflection coefficient at an interface (see Fig.7). Indeed, when irradiating a resonant medium, like a vapor, at the interface with some transparent window, the resonant change of refractive index induces a modification of the reflection, as predicted by the Fresnel formulae. Under *normal incidence*, the reflection coefficient R (in intensity) obeys :

$$R(\delta) = \left| \frac{n_w - n_v(\delta)}{n_w + n_v(\delta)} \right|^2 \tag{10}$$

with $n_w$ the refractive index of the dense window, and $n_v(\delta)$ the resonant index of the vapor, as evaluated for a frequency detuning $\delta$ relatively to the atomic resonance. Assuming $|n_v(\delta) - 1| \ll 1$, one calculates at first order the resonant reflectivity change $\Delta R(\delta)$ :

$$\Delta R(\delta) = -\frac{4n_w(n_w-1) \Re[(n_v(\delta)-1]}{(n_w+1)^3} \tag{11}$$

Equation (11) shows that the SR lineshapes is governed by the dispersion (or refractive index) of the vapor, and is independent of the vapor absorption (within the first order approximation). This is why SR spectroscopy has been used as a way to probe an optically opaque vapor: as opposed to absorption or fluorescence spectroscopy, the SR signal, originating from the reflection *at the interface*, is insensitive to the cell length. As discussed below, the actual typical distance of observation is on the order of an optical wavelength; this point can appear somehow hindered by the assumption -as in Eq. (10)- of spatial homogeneity of the refractive index.

In a microscopic description of a SR experiment, the field reflected at the interface $\mathbf{E}_r(\delta)$ is the sum of the field $\mathbf{E}_r^0$ reflected from the dielectric interface, and the field reflected by the vapor $\Delta \mathbf{E}_r(\delta)$. $\Delta \mathbf{E}_r(\delta)$ results from the coherent summing of the field emitted by all oscillating dipoles, driven by the incident field in the resonant vapor $\Delta \mathbf{E}_r(\delta)$. Under the standard assumption $|\Delta \mathbf{E}_r(\delta)| \ll |\mathbf{E}_r^0|$, which is analogous to the one used to derivate Eq. (11), one gets:

$$\Delta R(\delta) = \frac{2\mathbf{E}_r^0 \Re[\Delta \mathbf{E}_r(\delta)]}{|\mathbf{E}_r^0|^2} \tag{12}$$

In Eq.(12), we have conventionally assumed an optical phase such that $\mathbf{E}_r^0$ is real; also, for purposes of simplicity, all optical fields and oscillating dipoles are expressed in the rotating frame. Note that in more complex geometries (such as a multilayer window, possibly including an absorbing layer -*e.g.* metallic coating-), $\Delta R(\delta)$ may become proportional to a complex admixture of $\Re e[\Delta \mathbf{E}_r(\delta)]$ and $\Im m[\Delta \mathbf{E}_r(\delta)]$, with heavy consequences on the predicted lineshapes (Chevrollier *et al*., 2001).

In the calculation of the resonant reflected field $\Delta \mathbf{E}_r(\delta)$, one has to sum up the field radiated by the atom dipoles, that are spatially distributed in the vapor. Hence, a phase factor *exp*(ikz) (with k = 2π/λ the wave number of the incident radiation) has to be taken into account for the propagation of the incident field to the locally induced dipole $\mathbf{p}$(z) (see Fig.7). A similar additional phase contribution appears for the coherent summing-up – at the interface z = 0 – of the backward radiated field. One gets:

$$\Delta \mathbf{E}_r(\delta) = \frac{1}{2i\varepsilon_0 k} \int_0^\infty \mathbf{p}(z,\delta)\, exp(2ikz)\, dz \qquad (13)$$

In Eq.(13), the phase modulation factor *exp*(2ikz), analogous to a mismatch factor in nonlinear optics, is responsible for the spatial resolution of the SR method, defining a "coherence length" (*e.g.* λ/4π), on the order of a reduced wavelength $\lambdabar = \lambda/2\pi$. This confirms that, when the medium response is spatially homogeneous (*i.e.* $\mathbf{p}$(z) = $\mathbf{p}$), increasing the length L (L>>λ) does not increase the reflected field. An interesting consequence of this short coherence length is that the SR signal can be dominant in the spatial regions where $\mathbf{p}$(z) varies sharply (*i.e.* on a λ/4π scale). This is the reason for SR to appear as a convenient method for probing the surface interaction : in an homogeneous medium, only the vicinity with the surface is susceptible to induce rapid changes of $\mathbf{p}$(z). Note at last that in Eq.(13), it has been essential to assume that the density of the resonant medium remains low enough, so that the propagation is unchanged by resonance (*i.e.* k remains real and identical to its value in vacuum).

### 3.2.2 *Atomic response and atomic motion*

#### 3.2.2.1 *The resonant atomic response*

For a vapor of motionless atoms, and neglecting the surface interaction effects, the induced dipole $\mathbf{p}$(z) is usually derived from a simple *linear* resonant response such as:

$$\mathbf{p}(z, \delta) = \mathbf{p}(\delta) = \mathbf{p}_0 \frac{\gamma/2}{\delta - i\,\gamma/2} \tag{14}$$

with $\gamma$ the relaxation rate of the so-assumed Lorentzian resonance, and $i\mathbf{p}_0$ the dipole response at resonance; note that in Eq. (14), $\mathbf{p}_0$ is real, so that on resonance ($\delta = 0$), the induced dipole oscillates in quadrature with the exciting field, and radiates a field whose phase is opposite to the one of the incident field, signing the absorption process. It is hence easily verified from Eqs.(12-14) that for a homogeneous semi-infinite vapor, the SR signal is purely governed by the dispersive response of the atomic dipole, justifying the elementary approach of Eq.(10).

In the early descriptions of the SR spectroscopy, the effect of the atomic motion was simply described as an inhomogeneous broadening on the resonance frequency, with the detuning $\delta$ replaced by a Doppler-shifted detuning, *e.g.* ($\delta$-kv$_z$) with v$_z$ the normal velocity for SR under normal incidence. This approach predicts for the SR lineshape a Doppler-broadened dispersion (or more generally, a dispersive Voigt profile, see Fig.8). It is actually valid only when the Doppler broadening is small relatively to the homogeneous broadening $\gamma$. Early high-resolution experiments performed by Cojan (1954), and confirmed in the laser spectroscopy era by Woerdman and Schuurmans (1975), have established that the lineshape is narrower than predicted by this approach. Indeed, the steady-state response given by Eq.(14) does not describe the response of atoms departing from the wall, which is a *transient* build-up response: as long as an atom is on the wall, its energy structure is considerably perturbed, and it is insensitive to the nearly resonant excitation of the incident field. At least two notable consequences can be derived from this behavior:

(i) the response in SR spectroscopy is essentially "nonlocal", as due to the transient effects and to the atomic motion  (Schuurmans, 1976). Indeed, the atomic response of a given velocity group $\mathbf{p}(z,v_z)$ depends not only on the incident field $\mathbf{E}_0(z)$, but also on the field $\mathbf{E}_0(z')$ for all the z' values included in the (classical) trajectory explored before reaching z.

(ii) The *exp* (2ikz) modulation factor tends to scramble the contribution of atoms exploring many wavelengths. This enhances relatively the contribution of those atoms that are slow enough (along the *z*-axis) to reach their steady-state response within $\sim \lambdabar$  and it is at the origin of a sub-Doppler contribution in the SR lineshape at normal incidence. It shows also that those atoms that spend a long time interacting with the surface bring an important contribution to the SR signal.

Note that in spite of the nonlocal atomic response, the evaluation of the SR signal derived from the Fresnel formulae (such as Eqs.(10-11) for the normal incidence) still stands, provided that the *local* refractive index is replaced by an "effective index". This index relies on an "effective susceptibility" of the medium (Nienhuis *et al.*, 1988), that spatially integrates all of the atomic response, and is no more intrinsic to the medium. It is notably dependent upon the angular orientation of the incident beam.

### 3.2.2.2 The SR lineshape and the sub-Doppler logarithmic singularity

The standard analysis of the SR lineshape -in the absence of surface interaction- was developed by Schuurmans (1976). It is valid in the case of a Doppler-broadened two-level atom system in the regime of a linear field-atom interaction (*i.e.* saturation effects are to be neglected). It essentially integrates *separately* the contribution of the atoms arriving *onto the surface*, assumed to be in a steady state of interaction with the resonant light field, and of the atoms *departing from the surface*, experiencing a transient regime evaluated as a function of the elapsed time $\tau = z / v_z$ since the departure from the wall.

One of the major results is that, simply assuming a symmetric velocity distribution, *these two separate contributions are identical.* The SR lineshape (see Fig.8) is hence the convolution of a two-level Lorentzian response with a *half-Maxwellian* (assuming a thermal velocity distribution). This coincidence between the two contributions, later shown to survive when a surface interaction is included in the model (Ducloy and Fichet, 1991), can be traced back to a compensation between the $2kv_z$ Doppler shift between the two velocity groups $v_z$ and $-v_z$, and the phase factor *exp* (2ikz) typical of SR spectroscopy.

An important property of this convolution is that, due to the asymptotically slow $v_z^{-1}$ dependence (for $v_z \rightarrow \infty$) of the dispersive Lorentzian in $[(\delta-kv_z)-i\gamma/2]^{-1}$, the convergence of the velocity integration is ensured only thanks to the finite tails of the velocity distribution. This means that, for a given detuning, the SR response cannot be seen as originating in the contribution of a single "velocity group", as it is the usual case in absorption (*i.e.* in the "large Doppler width approximation", that cannot be used here for a dispersive lineshape). Rather, the (weighted) contribution of all atomic velocities has to be included. The abrupt singularity ($v_z=0$) of the half-Maxwellian distribution is responsible for the specific narrow SR signal at line center ($\delta = 0$) whose amplitude would diverge logarithmically if the Doppler width would be asymptotically large. This narrow response, evolving like $ln\left(\dfrac{ku\,\gamma/2}{\delta^2+\gamma^2/4}\right)$ - with u the thermal velocity- appears superimposed to a broader dispersive Doppler-broadened profile (see Fig.8).

### 3.2.2.3 The narrowing of SR lineshape with FM modulation

This remarkable sub-Doppler feature of SR spectroscopy, that singles out the $v_z=0$ contribution, has been turned into an efficient Doppler-free method when the Lebedev group (Akusl'shin *et al.*,1982) recognized that the SR lineshape yields a Doppler-free signal once frequency-derived. Technically, such a derivation is conveniently performed with a FM (frequency-modulation) applied to the incident field, and demodulation of the SR signal.

The (FM) SR signal turns out to be a pure Doppler-free dispersive Lorentzian in the infinite Doppler width approximation, now allowed as the derivation ensures the velocity integration convergence. The selected velocity component ( $k.|v_z| < \gamma/2$ ) corresponds to atoms moving by less than $\lambda$ -normally to the surface- during the relaxation time $2\gamma^{-1}$ associated to the optical width.

### 3.2.3 SR lineshapes in the presence of an atom-surface interaction potential

#### 3.2.3.1 The general case

Under the assumption of a linear regime of atom-light interaction, and solving the problem of the nonlocal atomic response with the assumption of atomic linear trajectories at constant velocity, a formal calculation of the SR signal has been derived in the Ducloy and Fichet (1991) paper, that accounts for a z-dependent surface interaction potential [*i.e.* $\delta = \delta(z)$ = $\omega-\omega_0(z)$, with $\omega_0$ the z-dependent resonance frequency of the atom, and $\omega$ the irradiating frequency], also including a possible z-dependent transition width (*i.e.* $\gamma = \gamma(z)$). The calculation generally requires the evaluation of a triple integral, through (i) the averaging over the velocity distribution, (ii) the spatial averaging typical of the SR spectroscopy, and (iii) the spatio-temporal integration associated with the buildup of the oscillating atomic dipole.

The FM technique, shown above to emphasize the contribution of slow atoms, is particularly well-suited to the study of atom-surface interaction, because it enhances the contribution of those atoms interacting for a long time with the surface. Technically, the velocity integration is hence replaced by the sole response of the zero-velocity group -*i.e.* atoms slow enough to exhibit a null Doppler shift, although the non local response, even for slow atoms, still demand a double spatial integration- . For this simplification, it is sufficient to assume that the Doppler broadening is large compared to both the homogeneous width, and the surface interaction, as evaluated at a distance $\sim \lambda$ (note that a diverging interaction at small distances remains compatible with this reduction).

*3.2.3.2 The specific case of the $z^{-3}$ vW interaction*

The double integral of the general FM SR signal, or at least one of the step of integration, can sometimes be evaluated analytically. Apart from the simple case of an exponentially decaying potential –as induced by the dipole force associated to an additional evanescent field-, a (vW type) $z^{-3}$ potential has been shown (Ducloy and Fichet, 1991) to enable such an analytical integration. Hence, a single numerical integration permits to obtain universal (dimensionless) FM SR lineshapes in the presence of a vW interaction (see Fig.9 ). An essential point is that under the simple approximation of infinite Doppler width, all features of these calculated lineshapes can be traced back to a single dimensionless parameter:

$$A = 2 \, k \, \boldsymbol{C_3}/\gamma \qquad\qquad (15)$$

with $\boldsymbol{C_3}$ the coefficient of the $z^{-3}$ shift. Note that in this spectroscopic approach, $\boldsymbol{C_3}$ has been defined for the probed $|i> \rightarrow |j>$ transition, *i.e.* $\boldsymbol{C_3} = C_3(|j>) - C_3(|i>)$ and $\omega_0(z) = \omega_0 - \boldsymbol{C_3}\, z^{-3}$. The optical width is here assumed to be constant with the distance ($\gamma(z) = \gamma$), *i.e.* it ignores the effect of a possible resonant transfer (see sections 2.5 and 4.4).

For low values of A ( $0 < A << 1$), the vW interaction is essentially a perturbation that imposes a combined shift and distortion to the essentially dispersive Doppler-free lineshape. In the strong vW regime (A>1), the lineshape is so strongly modified that one cannot recognize the original antisymmetry predicted for FM SR with A = 0. Various shapes are obtained, evolving from a red-shifted absorption-like shape for moderate A values (*e.g.* in the A ~ 1-10 range), to multiple oscillations on the red side (for A $\geq$ 100). The calculation can also be performed for A < 0 (Fichet *et al.*, 1995 b), as it occurs when a vW repulsion is exerted onto the excited state (see section 2.4) or if, for some peculiar reasons, one has $C_3(|j>) < C_3(|i>)$. Remarkably, a red-shift of the SR lineshape is still predicted in this case, although the local atomic resonance is shifted to the blue. This seemingly paradoxical result originates in the particular spatial averaging typical of SR spectroscopy (see Chevrollier *et al.*, 1992, and notably the appendix). On the other hand, in spite of their closely resembling lineshapes in the perturbative regime $|A| <1$, red-shifting (A > 0) and blue-shifting (A < 0) vW interactions can be distinguished by their amplitudes, and by their apparent width (see Failache *et al.*, 2003).

*3.2.4 Beyond several simplifying approximations*

The set of universal (FM) SR lineshapes that, as will be seen in section 4, provides the basis for a large part of our experimental investigations, has implied several current approximations. It is worth discussing the major ones, in order to understand how experimental constraints may lead to a possible violation of these approximations.

*3.2.4.1 Absorption*

The validity of Eq.(13) assumes that there is no absorption, in particular because the *exp* (2ikz) factor assumes that in the resonant medium, the propagation does not differ from the propagation in vacuum. An expansion at first order in absorption (or, equivalently, in atomic density) is sometimes tractable, but already leads to the blue shift first predicted by Schuurmans (1976). Note that the observation of the SR signal, essentially related to a probe region of extension ~ $\lambda$ , often occurs in experimental conditions for which the absorption is non negligible on a wavelength scale. This shows that, in the principle, extrapolation of the SR data to the limiting situation of a weak absorption is always needed. The problem hence becomes extremely complex, with local field (Lorentz field) corrections (Maki *et al.*, 1991) to be taken into account, and most often a non-exponential attenuation of the irradiating field (Vartanyan *et al.*, 1995, see also Vartanyan and Weis, 2001).

*3.2.4.2 Symmetry of the velocity distribution*

The remarkable compensation appearing in the linear regime between the phase shift imposed by the transient response and the Doppler shift separating $v_z$ and $-v_z$ velocity groups, permits to calculate the SR lineshape with the sole contribution of the arriving atoms -or of the departing atoms- provided that the velocity distribution (over $v_z$ ) is symmetric (Ducloy and Fichet, 1991). Although this assumption looks reasonable, it may be not satisfied in various situations of experimental interest. For a gas sample submitted to an additional optical pumping, the atomic polarization -mostly induced on the *arriving* atoms- can be partly destroyed through a collision onto the wall, leading to an asymmetry in the velocity distribution for a given atomic state. More generally, it is actually fascinating to note that, for a gas at equilibrium, the common idea of an isotropic Maxwellian velocity distribution may become invalid close to a surface (Comsa and David, 1985). Indeed, various reasons at the microscopic level, including the structural details of the surface, angular properties of desorption, surface roughness, or quantum effects on the atomic trajectories, may actually contribute to invalidate the Lambert-type "*cos* θ " angular law for departing atoms, whose validity has been seldom tested (see however Grischkowski, 1980; Bordo and Rubahn, 1999). Note that the selection of "slow" atoms, typical of the FM SR method, is moreover a way of testing the large angular values (*i.e.* trajectories at a grazing incidence with respect to the window plane, or $\theta \sim \pi/2$), usually out of reach for most of the mechanical methods. Let us mention that thanks to a relatively simple experiment of *nonlinear* selective reflection, the contribution of the slow atoms was measured to be approximately twice smaller for the arriving atoms, than for the departing atoms (Rabi *et al.*, 1994). Unfortunately, it has not been possible, through these measurements, to attribute unambiguously this effect to a surface specificity of the velocity distribution, rather than to a differential saturation effect.

### 3.2.4.3 De-excitation at the arrival onto the surface

The SR theory assumes that arriving atoms get de-excited when hitting the surface. Hence, the contribution of departing atoms is evaluated as a transient excitation, built-up from the ground state, and function of $\tau = z/v_z$ (with $\tau$ the time elapsed since the atom has departed from the wall). Actually, it is conceivable that an atom, in a wall collision that would be "instantaneous" (*i.e.* on a time scale $\sim 10^{-12}$ s), does not relax all of its internal energy. However, the SR spectroscopy performs a kind of coherent measurement, sensitive to the oscillating atomic dipole, rather than to the atomic excitation. Hence, the tremendous interaction potential exerted by the surface at close distance, which is moreover state-dependent, guarantees (Ducloy, 1993) that the prepared coherent superposition of states is washed out in a surface collision process, even for an atom that could remain energetically excited in the collisional process.

### 3.2.4.4 Finite Doppler width

The FM SR is a genuine Doppler-free method only in the frame of the "large Doppler approximation", assuming $ku/\gamma \to \infty$, and also $\delta \ll ku$. In a realistic case, even for $ku/\gamma = 100$, and as due to the rather slow evolution of SR lineshape with the $ku/\gamma$ factor (the "divergence" being only logarithmic), the FM SR technique does not yield a signal totally independent of the velocity distribution, and some corrections must be applied to the pure dispersive Lorentzian model (Papageorgiou *et al*, 1994; Failache, 2003). In standard cases, these corrections affect only a contribution associated with the tails of the velocity distribution, with no essential effects on the predictions for the nearly Doppler-free SR resonance.

*3.2.4.5 Normal and oblique incidence*

The sub-Doppler contribution, turned into a genuine Doppler-free signal for FM SR, has been predicted under the hypothesis of an irradiation at normal incidence. The narrowing occurs because of an identity between the "Doppler axis" (along which the velocity is counted for the estimate of the Doppler shift), and the normal axis, along which the transient effects are evaluated. Under an oblique incidence, the FM SR Doppler-free lineshape is turned into a lineshape sensitive to a *"residual" Doppler broadening*, on the order of $ku\theta$, with $\theta$ the incidence angle of the irradiating beam in the vapor (Nienhuis *et al.*, 1988; Oriá *et al.*, 1991; Chevrollier *et al.,* 1991 and 1992). This broadening is negligible as long as $\theta << \gamma/ku$.

.

*3.2.4.6 Atomic trajectories: Spectroscopy vs. mechanical effects*

The essence of the SR lineshape model that we have described above consists in atomic trajectories traveled at a constant velocity. It neglects any possible curvature imposed by the normal force exerted by the atom-surface potential. Such an assumption may appear rather crude, especially with respect to the special contribution of the slowest atoms, that are the most sensitive to the mechanical effects of the potential. We have attempted to enhance such a modification of the velocity distribution close to the wall by studying (Papageorgiou, 1994 and Papageorgiou *et al.,* 1995a) the influence of an auxiliary blue-detuned repulsive evanescent wave on SR spectral lineshapes. However, the SR signal reveals sensitive not only to these possible changes in the atomic velocity, but also to the spectroscopic effect of the potential (Ducloy *et al.*, unpublished)

To discriminate between dominant mechanical or spectroscopic effects, the relative value of the Doppler shift associated to the potential-induced velocity change, and of the spectroscopic shift imposed by the potential, can provide a rule of thumb. For a "typical" step of the potential $\Delta U$ (the "step" being distributed over a distance comparable with the spatial resolution $\sim \lambdabar$ , notwithstanding a possible divergence of the potential very close to the wall), the velocity change $\Delta v$ is given by :

$$(v_0 + \Delta v)^2 - (v_0)^2 = 2\ \Delta U/m \tag{16}$$

with $v_0$ the initial velocity, and m the atomic mass. In the FM SR approach, the typical selection of slow atoms is a restriction to velocities $|v_z| \le \gamma/k$. With a specific interest to this upper boundary $v_0 = \gamma/k$, and assuming $\Delta v << v_0$ , one calculates from Eq. (16):

$$\Delta v =\ \Delta U\ k/\gamma m \tag{17}$$

According to our criterion, the Doppler shift associated to the mechanical effect dominates over the interaction potential if $\hbar k \Delta v \geq \Delta U$. In our approach, this appears to be independent of the potential $\Delta U$, and simplifies to:

$$\hbar k^2 \geq m\gamma \qquad (18)$$

This shows that for our typical experiments on a heavy atom like Cs at $\lambda \sim 1\mu m$, and for rather broad resonance lines ($\gamma \sim$ few $(2\pi)$ MHz), the spectroscopic effect is dominant. Conversely, the observation of the mechanical effects would demand a particularly light atom, along with a very narrow optical transition. More generally, the criterion defined by Eq.(18) implies that the mechanical effect dominates only when the selected atomic momentum $mv_0$ is smaller than the recoil effect $\hbar k$. Such a condition seems at odds with uncertainty principle: it shows convincingly that the mechanical effects of the potential should not be dealt with in the frame of classical mechanics.

### 3.3 *Nonlinear Selective Reflection*

The principle of probing an interface through SR spectroscopy can be extended from *linear* spectroscopy (*i.e.* with a single incident beam, whose intensity is assumed to be weak enough to avoid saturation) to *nonlinear spectroscopy*, when the medium is sensitive either to the saturation induced by the single irradiating beam, either irradiated with multiple beams. However, it should be recalled that one essential peculiarity of SR spectroscopy is the logarithmic enhancement of the slowest atoms *in the wall frame* (and the genuine velocity selection in the FM mode)*, and that this behavior is intimately connected to the *linear* atomic response.

### 3.3.1 *Saturation with a single irradiating beam*

Under a relatively strong resonant irradiation, in the same SR scheme as described in section 3.2.1, saturation of the optical transition can appear through optical pumping of the resonant atoms to a third level. The onset of such a pumping usually requires moderate intensities owing to the long relaxation time of the pumping (*e.g.* hyperfine optical pumping). For this reason, the arriving atoms - already in the steady-state - are much more sensitive to the optical pumping effects than the departing atoms. A related experimental situation has been analyzed by Vuletić *et al.* (1994): two different effective saturation intensities have been shown to appear, that were attributed to the different behavior of arriving and departing atoms, the arriving atoms being sensitive to saturation with an allowed hyperfine optical pumping, while the saturation for departing atoms is related with optical saturation of a pure two-level system.

### 3.3.2 *Multiple beams nonlinear SR spectroscopy*

In volume, nonlinear (NL) spectroscopy with multiple beams irradiation, such as pump-probe spectroscopy, is a very efficient way to impose the selection of a given velocity group. Even with a system as simple as a three-level system, a large variety of schemes can be found (ladder scheme, cascade, Λ-type, V-type,...). When one extends these techniques to NL SR spectroscopy at an interface, the variety of situations (see *e.g.* Schuller *et al.*, 1991, 1993, 1996) is, at least, as large because propagation from the window, or to the window (with unavoidable reflections) generally induce different behaviors. In addition, the overlap of various incident beams, can sometimes generate extra non phase-matched nonlinear emission from the bulk (Le Boiteux *et al.,* 1987; Amy-Klein *et al.*, 1995; Sautenkov *et al.*, 1997).

A specific point of NL SR spectroscopy is that, while it is easy for the arriving atoms to interact with the pump excitation (at least in the large region where the surface interaction is negligible), more complex features can appear in the pump interaction with departing atoms, owing to their transient behavior. This generally means that, in addition to narrow resonances associated to velocity groups arbitrarily selected by the NL excitation, as occuring in volume spectroscopy, one still predicts, in NL SR spectroscopy, a specific NL response of the slow atoms yielding, when the surface interaction is neglected, a signal centered on the probed transition (Rabi *et al.*, 1994 ; Gorris-Neveux *et al.*, 1995).

Up to now, the atom-surface interaction has been neglected in most of the specific theoretical approaches for NL SR. Indeed, the calculations are most often heavy (even in the lowest order limit, *i.e.* third order) because the transient behavior of a NL polarization should be evaluated, and then spatially integrated. Typically, the triple integral of linear SR mentioned in section 3.2.3 has to be replaced at least by a quintuple integral, with no obvious simplification. Another point worth mentioning is that the integrand of the NL SR response, equivalent to the $\mathbf{p}(z)\,exp\,(2ikz)$ term appearing in Eq.(13), include various spatial frequencies, as a consequence of the multiple beams NL excitation. This makes the depth of the coherently probed region difficult to estimate. This may explain why no surface interaction effect has been clearly identified in experiments based upon NL SR spectroscopy, even when the excited atomic levels are supposed to undergo a strong vW response (Gorris-Neveux *et al.*, 1995). However, when a strong spatial dispersion is induced (van Kampen *et* *al*., 1998), notably when the pumped region is confined close to the interface (*e.g.* pumping with a confined or evanescent wave), some specific possibilities for the probing of the atom-surface interaction may exist. Also, the possibility of selecting an arbitrary velocity group close to the surface may provide quantitative indications on the actual velocity distribution close to the surface (see sub-section 3.2.4.2),  with a possible differentiating between the slowly arriving and slowly departing atoms (Rabi *et al.*, 1994).

### 3.3.3 *Pseudo-thermal pumping in an excited state*

A convenient situation combines the *linear probing* on a transition between excited states, with the generation of an  artificial (quasi-) thermal population, as induced thanks to an auxiliary pumping scheme. It permits indeed to benefit from the well-understood  knowledge of linear SR spectroscopy, while reaching excited atomic levels by a stepwise process. Such a technique has been notably implemented in an experiment described in the section 4 (Failache *et al.*, 1999 and 2003), with a broadband pumping, in an off-axis geometry (*i.e.* yielding Doppler-broadened pumping). Note that it may be not obvious that the population of departing and arriving atoms are equal very close to the surface, because the pumping of the departing atoms occurs in a transient regime of interaction.

### 3.4 *Evanescent wave spectroscopy*

The general technique of Attenuated Total Reflection (ATR) consists of the coupling of an inhomogeneous field  to a resonant  medium : although an inhomogeneous field does not propagate energy, an energy transfer between the field and the medium can occur. This transfer can be detected on a propagating field coupled to the inhomogeneous one. In an elementary scheme, the evanescent wave (EW) spectroscopy  set-up (see Fig.10) consists of a traveling wave entering into a prism with an internal incidence angle exceeding the critical angle for total reflection: no traveling field  emerges out in the vacuum or dilute resonant medium. On resonance, the reflected field is however attenuated (Carniglia *et al.,* 1972; Boissel and Kerhervé, 1981). Analogous ATR observations are expected when the inhomogeneous  field  is the field  of a surface plasmon, provided in a geometry like the one proposed by Kretschmann and Raether (1968).

Inhomogeneous fields are intrinsically confined close to the surface, with an amplitude exponentially decaying with the distance to the surface. In the principle, this makes EW spectroscopy - and generalizations- a suitable technique for the probing of an atom interacting with the surface. An essential difference with SR spectroscopy at an interface under a *real* incidence angle is that in the EW technique, the incident field in the vapor can be seen as propagating under an *imaginary* incidence: hence, the EW technique is essentially sensitive to absorption processes, rather than to the dispersion (Simoneau *et al.*, 1986).

Another difference with SR spectroscopy under normal incidence appears when atomic motion is considered : indeed, the Doppler shift in EW spectroscopy is counted along the velocity component parallel to the phase propagation (*i.e.* parallel to the interface) while the transient atomic response, intrinsic to the spatial inhomogeneity of the EW field, is counted with the motion along the normal to the window. Hence, while in regular SR spectroscopy, the transient behavior of the atoms helps to select the slowest atoms, EW spectroscopy is plagued both with transit time thermal broadening, and with Doppler broadening.

Extending to the EW techniques the well-known volume Doppler-free saturated absorption, we had developed a Doppler-free EW spectroscopy (Simoneau *et al.*, 1986), in which the pump and probe fields are evanescent fields induced with light beams that counterpropagate in the prism. However, the finite time spent in the evanescent fields strongly alters the efficiency of the velocity selection, so that the optimal spectroscopic resolution is achieved  only when the evanescent  fields  have a large spatial extension. In this case, Doppler-free EW spectroscopy is only weakly sensitive to the region  where long-range atom-surface interaction is important, limiting applications involving the monitoring of atom-surface interaction. However, atomic residence time could be evaluated through a dephasing measurement in this EW technique (Bloch *et al.*, 1990 ; Oriá *et al.*, 1992; see also de Freitas *et al.*, 2002), establishing a possibility of probing with high resolution spectroscopy a phenomenon related to the *short-range* atom-surface interaction.

### 3.5  *Spectroscopy in a thin vapor film: Micro- and Nano-cells*

The narrow spectral features characterizing SR spectroscopy under normal incidence are not that much typical of a reflection process, but associated to the transient regime of interaction undergone by atoms located in the vicinity with the surface. Hence, an analogous enhancement of the slow atoms contribution is expected to appear in transmission spectroscopy, as was demonstrated by Briaudeau *et al.* (1996, 1999 and refs. therein). Practically, such an effect is observable, under normal incidence, when the vapor cell is not too long -relatively to an average distance traveled by an atom before reaching a steady-state of interaction-, and when the vapor remains dilute enough so that the atoms fly from wall to wall. This last condition implies that the atomic "mean" free path in the short vapor cell is anisotropic, and justifies an overweight of atoms with small normal velocities (see Fig. 11)

A major difference with SR spectroscopy is that, in transmission spectroscopy, there is no more the oscillating phase factor *exp* (2ikz) factor (see Eq.(13)), that has favored the detection of atom-surface interaction through the enhanced contribution of regions where the atomic response varies rapidly over λ. Rather, the signal originates from the whole cell. This explains why the first demonstrations of these novel sub-Doppler features in transmission spectroscopy, performed with micro-cells whose thickness spanned in the 10-1000 μm range, were unsensitive to atom-surface interaction. Conversely, and as will be discussed in section 5 with more details, the recent development of sub-micrometric vapor cells ("nano-cells") (Sarkisyan *et al.*, 2001) seems a promising tool to explore vW atom-surface interaction for a given range of distances to the surface, eventually much smaller than those currently reached with SR spectroscopy (Dutier *et al.*, 2004 a).

## 4.     SR spectroscopy as a diagnostics tool of the Atom-Surface interaction

This section describes how SR spectroscopy has been used to measure the vW atom-surface interaction. Starting from the most elementary observations, when the dielectric image coefficient of the surface can be considered as a constant, it describes how an effective measurement of the $C_3$ value is performed, yielding a variety of information on the atomic state parameter affecting the vW interaction. It is illustrated with some of the effects associated with the resonant coupling between the atomic excitation and the surface modes. It ends up with a search for anisotropic effect in the vW interaction.

### 4.1 *Elementary observation of the vW interaction in linear SR spectroscopy*

#### 4.1.1 *Experimental set-up*

The observation of a SR spectroscopy signal basically requires a resonant narrow-linewidth tunable laser, a vapor cell enclosed in a container with at least one transparent window (wedged whenever possible), and a sensitive low-noise detector, in order to monitor conveniently the weak resonant change in the reflection coefficient, relatively to a non resonant reflected background on the order of several percents of the incident intensity. An auxiliary reference set-up, providing a signature of the volume resonance such as obtained with a saturated absorption (SA) set-up, is at least convenient to monitor the effect of the vW atom-surface interaction on the SR signal. Also, due to the higher resolution provided by the FM SR technique, a FM -of a small amplitude- is often applied, either to the irradiating beam with an external modulator, either to the laser itself, notably when a semi-conductor laser is used, for which it is easy to perform a FM with a modulation of the drive current. The FM SR signal is acquired after processing of the photodetector signal through a phase-sensitive lock-in detection. Note that, relatively to an *a posteriori* frequency-derivation of the recorded SR lineshapes, the FM technique eliminates the d.c. noise (and low-frequency noise) of the laser source.

### 4.1.2. *Typical observation in linear SR*

In volume spectroscopy, the weakness of an absorption line is often compensated for by an increase in the absorption length, if not by a multi-pass scheme. In SR spectroscopy, the signal amplitude is typically comparable with the one expected for absorption on a depth as small as $\lambda$, and only an increase in the atomic density can enhance the signal magnitude above the sensitivity threshold. However, the atomic density cannot be increased too much because of the self-broadening of atomic lines, so that SR spectroscopy can hardly be applied to very weak lines - *e.g.* we are not aware of SR laser spectroscopy applied to molecular lines-. Also, discriminating between pressure effects -related to (volume) atom-atom collisions -, and surface effects is a major concern for most SR measurements.

Nearly all experiments in SR laser spectroscopy have dealt with alkali vapors, whose atomic density varies quickly with temperature. We provide here some numerical indications for the Cs $D_2$ line (852 nm), although the Na $D_2$ line, the first one to be studied (Woerdman and Schuurmans, 1975), exhibits a comparable behavior, but for a smaller vW interaction. A density on the order of $\sim 10^{13}$ at/cm$^3$ (*i.e.* T$\sim$100°C for Cs) allows a detectable reflectivity change on the order of $10^{-4}$, -the precise values depend on the considered hyperfine components, and on the optical properties of the reflecting window - , while the pressure self-broadening ( $\sim 10^{-7}$ Hz/at.cm$^{-3}$ ) remains negligible relatively to the natural width ($\sim 5$ MHz) (see Papageorgiou *et al.*, 1994 a). On the SR lineshapes, one recognizes (see Fig.12), superimposed to the Doppler-broadened dispersive wings, narrower peaks associated with the respective hyperfine components, that are hence partially resolved. In the FM mode and under normal incidence, one observes well-resolved Doppler-free resonances that are close to dispersive lineshapes but for a slight asymmetry, already noted by Akusl'shin *et al.* (1982). In addition, high-resolution spectra most often reveal a small red-shift relatively to a reference SA spectrum, but its physical origin had been ignored until our works. This shift, on the order of 2-3 MHz, is only a fraction of the total width (most often in the 10-30 MHz range, depending upon the experimental conditions, and in excess of the 5 MHz natural width). As discussed in the next subsection, this combined distorted dispersive lineshape, and apparent lineshift, can be traced back to the vW atom-surface interaction (Oriá *et al.*, 1991; Chevrollier *et al.*, 1992). Conversely, the sole observation of an apparent frequency shift, while neglecting the distortion, cannot permit to evaluate, even approximately, the vW interaction.

At higher atomic densities, the broadening of the SR lineshapes on resonance lines goes along with an increased (anti-)symmetry of the dispersive lineshape, that has even provided particular opportunities for laser frequency stabilization purposes (Ito *et al.*, 1991; Li *et al.*, 1998). This regime of density can provide an essential set of data to analyze the pressure broadening effects. It has also provide a way to study the Lorentz correction associated with local field effects (Maki *et al.*, 1991). For this high density regime, the vW interaction can be taken into account as an extra-correction term (Guo *et al.*, 1996: Ping-Wang *et al.*, 1997).

Observing transitions weaker than the strong resonance lines of alkali is also possible, although more difficult. One of the narrowest FM SR lines that we have once observed has been recorded on the 791 nm intercombination line ($^1S_0$-$^3P_1$) of Ba, although the very high operating temperature -in excess of 700°C- rapidly destroyed the sapphire window, and imposed a notable broadening (~1-2 MHz) to a very narrow natural linewidth (100 kHz) (Failache *et al*, unpublished). In the principle, such a narrow resonance should permit to select particularly slow atoms, and could be favorable to observe mechanical effects induced by the atom-surface interaction potential (see section 3.2.4.6). The second and higher resonance lines of alkali vapor exhibit weaker oscillator strengths, orders of magnitude smaller than the first resonance line: however, as discussed in section 2.3, the corresponding excited levels are much more polarizable and more sensitive to the vW interaction than the first excited state. In spite of the rather high atomic density usually required for the SR signal to be observable, and its correlated pressure broadening, the SR and FM SR spectra clearly exhibit special features that are the signature of the vW interaction in the "strong" regime (see 3.2.3.2.) (*e.g.* (FM) SR lines that are resembling a shifted absorption-lineshape, or an inverted dispersion, ..).

## 4.2 *The method of experimental measurement of the $C_3$ coefficient*

### 4.2.1 _Fitting method_

As discussed above (section 3.2.3.2), the FM SR lineshapes for a 1-photon transition much narrower than the Doppler width have been described by a family of universal (dimensionless) lineshapes depending upon a single-parameter A (see Eq.(15), and Ducloy and Fichet, 1991). For an isolated narrow FM SR experimental resonance, it is not difficult to evaluate if a given value of the A parameter (or a given range $A_m < A < A_n$) is able to describe the observed lineshape. From an experimental determination of the width $\gamma$, that is *a priori* dependent on the A value (*i.e.* $\gamma = \gamma(A)$), one estimates the relevant value of $C_3$. The search for an optimal fitting (Papageorgiou, 1994), performed with a least square fit method, involves homothetic factors between the experimental spectrum and the dimensionless model, and offset factors to locate the resonance frequency. Note that the fitting cannot allow a continuous change of the A parameter, with respect to the non-analytic dependence of the A-lineshapes. Rather, the amount of the minimal error $\varepsilon(A_i)$, found when optimizing for each value of a set of $A_1, ..., A_p$ parameters, provides a criterion to find the acceptable range of $[A_m, A_n]$ values, easily converted into a range of acceptable values of $C_3$. On this basis, extensive improvements, yielding a high reliability (see Fig.13) and consistency, have been developed along the years. They are detailed in the work of Failache *et al.* (2003).

### 4.2.2 _Checking the consistency of a vW determination_

In various situations, it happens that for a given SR lineshape, the accuracy on the $C_3$ value remains low. Aside from an insufficient sensitivity in the recorded spectra, it can happen, in the weak vW regime (A << 1), that a large range of A value seems acceptable for the lineshape, while the $\gamma$ value remains essentially governed by the peak-to-peak width of the quasi-dispersive lineshape. Two very different vW regimes can also exhibit seemingly analogous lineshapes. This is why the determination of the $C_3$ value is generally secured by a consistency check, comparing the lineshapes obtained under various pressure conditions (Chevrollier *et al.*, 1991 and 1992; Failache *et al.*,1999 and 2003). Varying in a controlled manner the $\gamma$ value (*e.g.* by density broadening), leads to phenomenological changes for the SR lineshapes, that become very remarkable in the strong vW regime (A >> 1). In spite of the observed changes induced by (volume) atom-atom interaction, the atom-surface interaction has to remain unchanged. Such a test, illustrated in Fig. 14, is so sensitive that it has permitted to simultaneously evaluate the coefficients governing the vW interaction, and the pressure shifts (Chevrollier *et al.*, 1991 and 1992). It also provides an adequate citerion to choose between two very different evaluations of the A value (Failache, 1999; Failache *et al.*, 2003).

### 4.2.3. *Comparing the $C_3$ experimental values with the theoretical predictions for SR experiments at the interface with a non dispersive material*

With the above fitting techniques, the $C_3$ value for a given atomic transition and material at the interface, has been determined with an uncertainty in a 10-30 % typical range. Note that this uncertainty, if partly statistical and noise related, also originates from a systematic biasing appearing on the extrapolated values for $C_3$ , when attempting to improve the modeling of the SR lineshape. Several series of measurements have been performed on the Cs $D_2$ line ($6S_{1/2}$-$6P_{3/2}$), that yielded $C_3 \approx 2$ kHz.$\mu m^3$ at a fused quartz interface (Oriá *et al.*, 1991; Chevrollier *et al.*, 1992 ; Papageorgiou *et al.*, 1994 a). With respect to the dielectric response coefficient of fused quartz ($\sim 0.35$ for all the virtual transitions relevant for the vW interaction exerted onto the $6S_{1/2}$ and $6P_{3/2}$ levels), there is an agreement, on the order of 30%, between the (slighlty higher) experimental and  the predicted theoretical value (Chevrollier *et al.*, 1992 ). Note also that the $C_3$ contribution of the excited state $6P_{3/2}$ is about twice larger than the one of the ground state $6S_{1/2}$ (with $C_3$ = $C_3(6P_{1/2})$ - $C_3(6S_{1/2})$). Comparable values are predicted for other resonance lines of alkali vapors, that are roughly confirmed by the experimental results, notably on the $D_1$ line of Cs, and on the D lines of Rb (see *e.g.* Gorris-Neveux *et al.*, 1997; Ping-Wang *et al.*, 1997).

For the weaker transition to the more excited Cs($7P_{3/2}$) level (reached with the second resonance line, $\lambda = 455$ nm), the evaluation of the $C_3$ value has required the removal of the pressure shift effects. It was been found to be an order of magnitude larger (Chevrollier *et al.*, 1991 and 1992) than for the $D_2$ line. This $C_3$ value originates essentially from the vW interaction exerted onto the excited level Cs($7P_{3/2}$). Although not performed systematically, a comparison between SR spectroscopy with a fused quartz window, and with a sapphire window, has evidenced a stronger interaction of the Cs($7P_{3/2}$) atoms with sapphire, as can be expected from a comparison of  the $[(\varepsilon - 1) / (\varepsilon + 1)]$ factor. The systematic experimental evaluation of the $C_3$ value yielded also a reasonable agreement with the theoretical prediction, although this one is also found  slightly below ($\sim 30$ %) (note that the early given experimental error bar $\sim$ 15  % did not account for some possible systematic errors). Note that the relevant virtual transitions for the Cs($7P_{3/2}$) lie in the far IR region, and that even in the absence of a resonant coupling of the atom excitation with a surface mode, the theoretical evaluation of the dielectric image coefficient may require slight re-evaluation relatively to the early and simplified estimates.

Recent preliminary experiments on the even weaker transition to the more excited Cs($8P_{3/2}$) level *via* the third resonance line of Cs at $\lambda = 388$ nm (Dutier *et al.*, 2004 a; Hamdi *et al.*, 2004), shows that, with a refined fitting method, a 10 % accuracy level on the $C_3$ determination is feasible on this line, hence requiring a careful control of the frequency scan.

### 4.3 *Observation of the resonant long-range coupling between a surface mode and an excited atomic level*

#### 4.3.1. *Predicting a resonant atom-surface coupling*

As already mentioned after Eq.(8), a strong enhancement (positive or negative) of a given virtual emission contribution to the $C_3$ value can occur for some materials, as the values of $\Re\mathrm{e}\left(\varepsilon(\omega_{ij}) - 1/\varepsilon(\omega_{ij}) + 1\right)$ are not limited (see Fig. 4). These resonances for this enhanced contribution are predicted to occur for the complex frequency poles of $[\varepsilon(\omega_{ij}) + 1]$. For such poles, the material is no more an optical "window", but is extremely absorbing (absorption typically occurs on one wavelength), while the resonance sharpness strongly depends on the refractive index in this spectral region. Practically, the evaluation of the surface dielectric response r($\omega_{ij}$) is extrapolated from tabulated data for $\varepsilon(\omega_{ij})$ (see *e.g.* Palik *et al.*, 1985), or from a fitting analytical model for $\varepsilon$., based upon a 3- or 4- parameters model for each bulk resonance. However, the uncertainty induced by this extrapolation can be notable in the resonance regions. Moreover, there remains some uncertainties about the precise resonant behavior, as due to possible differences (dopants or impurities, temperature, ...) between the actual window, and the samples used in the literature

To illustrate the possibility of a resonance in the vW interaction, we have notably concentrated on the virtual coupling between Cs(6D) and Cs(7P) : our first demonstration of a strong vW regime in SR spectroscopy, achieved on Cs (7P), had relied indeed on the strong coupling to Cs (6D) (virtual *absorption* falling in the 12-15 µm range). Sapphire, whose main bulk absorption resonances are located at ~15µm and ~20 µm, exhibits an isolated strong *surface* resonance (surface-polariton) across the 12 µm region with a quality factor Q ~100 (Fichet *et al.*, 1995). The real part of the surface response, associated to the vW shift, $\Re\mathrm{e}\left[\left(\varepsilon(\omega_{ij})-1\right)/\left(\varepsilon(\omega_{ij})+1\right)\right]$ exhibits a dispersion-like frequency response, while the imaginary contribution $\Im\mathrm{m}\left[\left(\varepsilon(\omega_{ij})-1\right)/\left(\varepsilon(\omega_{ij})+1\right)\right]$, that governs the surface-induced atomic level transfer (section 2.5, and below, 4.4) exhibits an absorption-like dependence. Note also that the exact position of these resonances depends on the sapphire birefringence axis orientation (see Gorza *et al.*, 2001, and Fig.6). With a *c*-axis perpendicular to the window ($c\perp$), the vW contribution to the shift associated to the 12.15 μm emission Cs($6D_{3/2}\rightarrow7P_{1/2}$) is predicted to be enhanced by a factor ~ -15 relatively to an ideal reflector. This contribution, that would contribute to ~ + 7 kHz.μm$^3$ over a total of 25 kHz.μm$^3$ for Cs($6D_{3/2}$) in front of an ideal reflector, induces a strong vW repulsion in front of a $c\perp$ sapphire window, that dominates over all other and non resonant contributions (see Table 1). It stands indeed for ~ -108 kHz.μm$^3$ in a total predicted value of ~ -96 kHz.μm$^3$ (see Failache *et al.*, 2003). For sapphire with a *c*-axis parallel to the window ($c//$), there is no more a strong repulsion, because the atomic resonance falls close to the c/f of the region of anomalous dispersion for $\Re\mathrm{e}\left[\left(\varepsilon(\omega_{ij})-1\right)/\left(\varepsilon(\omega_{ij})+1\right)\right]$: moreover, the detailed prediction (*i.e.* weak repulsion, or weak attraction) is highly sensitive to the sapphire resonance modeling.

The above predicted resonance behavior, that has led to the experimental demonstration of vW repulsion (Failache *et al.*, 1999 and 2003), relies on a specific coincidence between Cs($6D_{3/2}\rightarrow7P_{1/2}$) and $c\perp$ sapphire surface resonance. Actually such a coincidence is not so unusual. Indeed, for a high-lying excited atom, the most important virtual transitions (for the vW interaction) falls in the relatively far IR range, while transparency of the window is required in an energy range which is governed by a high atomic excitation. Moreover, the low quality factor of the surface resonance - *i.e.* $Q \leq 100$, relatively to the $Q \sim 10^8$ for atomic transition- makes it easy to find an atomic transition lying in the repulsive dispersive wing of a resonance. Such a standard material as YAG illustrates (see Fig.4) how easy it is to obtain a resonant behavior in the vW atom-surface interaction. It exhibits indeed numerous resonances in the 10-20 μm region in spite of a transparency region spanning from the UV to ~ 5 μm. This justifies that YAG also has been found to be repulsive for Cs($6D_{3/2}$) (Failache *et al.*, 2003).

.

### 4.3.2. *Measuring the atom-surface interaction in resonant situations*

The measurement of the vW surface interaction, even in the presence of a resonant coupling between the atom excitation and a surface polariton mode, is not intrinsically different from those measurements performed through SR spectroscopy from the ground state. However, the need to reach high-lying atomic states may require a multi-photon excitation. With respect to difficulties in the interpretations of NL SR spectra, -for which the surface interaction has never been included in the theoretical modeling, see section 3.3.2. -, the experiments have rather relied on a stepwise quasi-thermal pumping of the resonant level, followed by linear SR spectroscopy between the resonant level and the high-lying state.

The production of a quasi-thermal population in the resonant level requires a strong pumping and high atomic densities. This has made crucial an independent measurement of the collisional effect (broadening and shift) and provides some extra-difficulties in the evaluation of the "quasi-thermal" distribution. In spite of these difficulties, we could compare the material influence on the vW shift for several atomic systems (notably $Cs(6D_{3/2})$, $Cs(9S_{1/2})$, $Rb(6D)$) that are sensitive to a resonance with a surface mode (Failache 1999, Failache *et al,* 2003). The experimental accuracy can be on the order of 30 %. These experimental determinations themselves are in agreement with the theoretical modeling. Note that the hypotheses of a linear atomic trajectories still holds for all the regions significantly contributing to the SR spectrum : for $Cs(6D_{3/2})$ atoms with $v_z = 30$ ms$^{-1}$ , located in front of a $c_\perp$ sapphire window. the classical turning point is indeed located at $z \sim 10$ nm, a position that would imply a tremendous vW shift ~100 GHz.

### 4.4 *Förster-like energy transfer induced by the near-field coupling to the surface*

The resonant near-field atom-surface coupling affects not only the energy of atomic states through virtual processes, but it can induce a real change in the internal state, induced by the vicinity with the surface. In this energy transfer with a surface, that can be seen as the analogous of the Förster internal energy transfer between two distant molecules, an excited atom located in evanescent tail of a *surface polariton* mode, loses energy through a near-field transfer to the surface polariton. As already discussed in section 2.5, such an effect is governed by the same $z^{-3}$ law as the vW effect, and implies that the branching ratios of the excited state are strongly dependent on the distance to the surface. It dramatically affects the behavior of an atom in a high-lying state on its route towards a surface. Moreover, noting that resonant behaviors are actually quite common for highly excited atoms (see section 4.3.2), such a surface-induced energy-level transfer can appear as nearly universal, with the strength of the resonance essentially governing the distance from the surface at which this remote process occurs. Note that an analogous internal energy transfer, but induced with *surface plasmon* modes, had been discussed for embedded "atomic" species (namely, electronically excited molecules), but its efficiency was limited to a much shorter distance range, owing to the ultraviolet nature of the considered resonances (see *e.g.* the review by Chance *et al.*, 1978).

   The experimental demonstration of such a resonant transfer has remained qualitative, and was conducted with a comparison of different materials, at a distance to the surface controlled by the spatial resolution of a SR spectroscopy method. A SR spectroscopy experiment was conducted on the Cs $7P_{1/2} \rightarrow 10D_{3/2}$ transition ($\lambda = 1.298$ µm), in the presence of a strong (stepwise) two-photon pumping to $Cs(6D_{3/2})$. This transition is normally transparent, because the relevant levels are unpopulated, but by the surface-induced transfer from $Cs(6D_{3/2})$ to $Cs(7P_{1/2})$ (the spontaneous emission, falling at 12.15 µm, yielding a negligible contribution). The experiments (Failache *et al.*, 2002) evidenced  a strong difference between a (nonresonant) fused quartz window (no SR signal), and the resonant situation provided with a sapphire window (*c//*)  or a YAG window (observable SR signal). The surface-induced transfer has been estimated to occur for $z \leq 450$ nm for sapphire and $z \leq 300$ nm for YAG, and $z \leq 30$ nm for fused quartz. The spatial resolution of the SR diagnostics explains the observed differences. Such a difference is best seen for rather low Cs pressure, because at higher densities, energy-pooling collisions can compete with the surface-induced process, and induce a Cs $(7P_{1/2})$ population in the vapor volume.

### 4.5. *A search for anisotropy in the vW interaction: studying Zeeman components*

   Due to a sum rule, it has been shown (Chevrollier *et al.*, 1992) that the value of the $C_3$ ($|$ i, F$>$) $\rightarrow |$ j, F'$>$) coefficient for the scalar part of the vW interaction should be the same for all F $\rightarrow$ F' components as long as the degenerate $|$F, $m_F>$ levels are identically populated. Conversely, if the individual  Zeeman components are resolved, non negligible differences in the strength of the vW interaction are predicted.

   We have attempted to observe these differences, performing measurements in the regime of intermediate magnetic field (Papageorgiou, 1994 ; Ducloy, 1994; Papageorgiou *et al.*, 1995 b), with the Zeeman degeneracy removed. To escape from the difficulty of comparing the individual $A_{m_F \rightarrow m'_F}$  values, due to the relatively large uncertainty affecting these measurements, we have implemented a differential measurement, comparing the surface effect when the magnetic field that imposes the Zeeman structure is parallel, or perpendicular, to the surface. The experimental approach (Fig. 15) used a cubic cell, with two perpendicular windows being irradiated in identical conditions (the irradiated spots being very close to each other, for the spatial homogeneity of the Zeeman effect). In these experiments, performed on the well-known Cs $D_2$ line, reproducible differences, of the predicted order of magnitude, were observed but with details that were not in agreement with the vW non scalar calculations. One possible interpretation of these discrepancies has relied on the simultaneous modifications of the radiative properties of an excited atom in front of a surface (see section 2.5), that could affect locally the optical width of the transitions, and hence subtly modify the lineshapes. These modifications, ignored in the vW modeling, are also anisotropic, with a strong dependence on the relative orientation of the emitting dipole (Lukoscz and Kunz, 1977 b).

## 5. New developments and prospects9

Among the various prospects that we present in this section, some are natural extensions of the works with optical techniques presented in sections 3 and 4. They include, extension from SR spectroscopy to nanocell spectroscopy, and the effects of a non-zero temperature environment. New directions, well-suited to study the atom interaction in a range of very small distance to the surface, are also presently explored, including the deflection of atomic beam with nanoslits technology, and the interaction of atoms with strongly curved surfaces, such as are nanobodies.

### 5.1 Towards the exploration of strong confinement to the surface through spectroscopy in a nanocell

#### 5.1.1 Present technology and thickness measurement

It recently became possible (Sarkisyan *et al.*, 2001) to fabricate Extremely Thin vapor Cells (ETC) compatible with vacuum sealing and heating, so that an alkali vapor, of a controllable atomic density, can be studied when imprisoned in a container whose thickness can be as small as ~20 nm. In the present state-of-the-art, two thick transparent windows, carefully polished with an excellent planeity, are contacted to a ring-shaped spacer (typical thickness ~ 300 nm), and glued at a high temperature (mineral gluing). The external atmospheric pressure induces a curvature on the windows, and usually the local cell thickness varies smoothly from near contact in the central region, to ~ 1 µm in the peripheral regions. The challenging point of the construction is that the cell can resist, with negligible deformations, to a strong heating (up to 350°C), enabling variations of the atomic density on a large range. Alternately, for a construction with thinner windows, inserting the ETC in a vacuum chamber (Sarkisyan *et al.*, 2003) provides an adjustable local spacing, through the control of the environmental pressure (between 0 and 1 atm).

The parallelism of the two windows is intrinsically excellent (typical window diameter ~ 20 mm), implying a Fabry-Perot behavior, at least for the two internal windows. On the one hand, this provides a convenient interferometric method to estimate the local cell thickness - with an accuracy currently reaching 5 nm -, on the other hand, a spectroscopic signal in such a cell is not the simple signal associated with transmission -or, in alternate schemes, with selective reflection- of a traveling wave: rather, it is a systematic combination of absorption and reflection signals (Dutier *et al.*, 2003 b).

### 5.1.2 *Observation of surface induced effects*

When comparing the spectra obtained for various local thickness of the ETC, notable lineshapes differences are predicted, even in the absence of a surface interaction, as due to the Fabry-Perot behavior, which mixes up transmission and reflection response: depending on the thickness, the observed lineshapes appear shifted, and with a variable asymmetry (Dutier *et al.*, 2003 a). However, on the Cs resonance line, one has observed notable red shifts and well-characterized lineshape distortions for a thickness typically below 100 nm (Dutier, 2003; Dutier *et al.*, 2004 a, b, and Dutier, in preparation). This behavior, easily observed with the FM technique (see Fig. 16), can no longer be traced back to the mixture of dispersive wings and absorption-like lineshapes. The frequency shift increases quickly with decreasing the thickness -roughly like $1/L^3$ (L, cell thickness)-, while the observed lineshapes have been found in good agreement with predictions that include the estimated vW interaction (as estimated from theory or from previous SR experiments).

From a more systematic comparison between the experimental lineshapes, and the family of theoretical models for various strengths of the vW interaction, it should be possible to measure effectively the vW interaction, as it was done with SR experiments (section 4.2). Hence, with the spatial resolution intrinsically offered by such nanocells, it should be possible to test the law of spatial dependence in $z^{-3}$ of the vW interaction more effectively than with the SR technique, that always averages over $\sim\lambda$ . Note that here, the effective vW potential results from an interaction with multiple electric images. Also, for each ETC thickness, a full set of $\boldsymbol{C_3}$-dependent lineshapes has to be calculated.

Experiments with nanocells have also been performed on more excited levels, such as Cs (6D) level, either after a pumping on the $D_1$ or $D_2$ line, either in two-photon schemes. We have already evidenced, after a nearly homogeneous pumping on the $D_2$ line, the very large vW attraction (several 10's of GHz) exerted onto the Cs ($6D_{5/2}$) level for small thickness in the 20-50 nm range *i.e.* for an atom-wall distance remaining always below 25 nm. These observations remain however very preliminary because unwanted effects, such as dynamic Stark shifts, are often present, as the experimental conditions often require intense beams for the signal not to be too small. Also, for these distances, surface roughness of the windows could be an issue.

### 5.1.3 *Possibility of a level crossing induced by a surface resonance*

Extrapolating at much shorter distances the $C_3(\,|i\rangle)\,z^{-3}$ behavior asymptotically demonstrated for long distances, opens in the principle the possibility of an $|i\rangle$ - $|j\rangle$ level crossing as long as $C_3(\,|i\rangle) \neq C_3(\,|j\rangle)$ . This can be illustrated with Cs (6D) and sapphire (see Fig. 17). The fine structure sublevels of ($6D_{3/2}$) and ($6D_{5/2}$) are predicted to behave differently in front of a sapphire surface: the ($6D_{3/2}$)$\rightarrow$ ($7P_{1/2}$) virtual emission at 12.15 μm falls in a resonance of the surface modes of sapphire, inducing a repulsive behavior (for $c_\perp$ sapphire), while the $6D_{5/2}\rightarrow$ $7P_{3/2}$ virtual emission at 14.6 μm is non resonant, and lets unchanged the regular and relatively weak vW attraction (see Fig.4). With the knowledge of the respective $C_3$ values (associated to the asymptotic $z^{-3}$ behavior at $z \rightarrow \infty$) for Cs($6D_{3/2}$) and Cs($6D_{5/2}$), a level crossing is hence expected for a distance to the wall $z \sim 5$ nm. This estimated distance is only marginally modified (Dutier, 2003) if one takes into account the local modifications affecting the wavelength of the relevant resonant virtual emissions, as induced by the vW shift. Conversely, the energy where this level crossing occurs is highly sensitive to the local details of the vW potential.

Looking for a signature of such a level-crossing -which could turn to be an anti-crossing, depending on the non diagonal term of the vW interaction Hamiltonian-, we have started to investigate an anomaly in the wings of the spectral lines reaching (in the free-space) the Cs($6D_{3/2}$) and Cs($6D_{5/2}$) levels. For such a purpose, spectroscopy in an ETC seems to offer much wider possibilities than SR or evanescent wave spectroscopy, as due to a spatial resolution limited only by the cell construction. The observation of such effects have been attempted, with the help of a widely tunable laser source (Yarovitski *et al.*, in preparation), operating across the whole fine structure ($\sim$ 1 THz) of Cs(6D) from the resonant Cs($6P_{3/2}$) level. However, for the shortest cell thickness, the vicinity between the two walls, and their imperfections at short distances, may modify the conditions ensuring a resonant coupling between atom excitation and the surface modes.

### 5.1.4 *Atom-atom interaction in a tightly confined medium*

A significant observation of the atom-surface interaction in a vapor requires to eliminate the effects of atom-atom interaction, such as pressure broadenings and shifts. These collisional effects themselves may depend on the cell thickness, a parameter that can be conveniently varied in ETC spectroscopy. Let us recall that the long-range atom-atom interaction, that scales like $r^{-6}$ (r : the interatomic distance), and whose integration over a half-space leads to the atom-surface vW interaction (see section 1), can be viewed as a reaction of an atom to the field induced by its own e.m. fluctuations as mediated by the perturber atom. It has been predicted that this interaction should be affected by confinement in a cavity, when the dimensions of the cavity are smaller than -or comparable to- the relevant wavelengths of the dipole fluctuations (Cho and Silbey, 1996; Cho, 1999; Boström *et al.*, 2002). Indeed, the fluctuating dipole field can interact with the perturber atom following various propagation paths that include reflection onto the confining surfaces. Until now, these theoretical predictions had remained untested, because there was no available experimental methods to explore such a collisional regime. It seems also that these predictions have remained limited to the elementary situation of two atoms located at fixed positions, not providing an estimate of the overall effect resulting from an integration over a distribution of positions and velocities. Clearly, these atom-atom interactions that are mediated by a surface should be considered in various devices such as atom chips, and their detailed understanding will require a sufficient knowledge of the more elementary "atom-surface" interaction.

### 5.2 *Thermal effects*

In numerous cases, the temperature of the surface remains small enough so that the thermal energy $k_BT$ (with $k_B$ the Boltzmann constant, and T the temperature) is much smaller than the involved energy of the relevant atomic transitions. However, high-lying energy levels are usually connected to neighboring levels that are not very distant in energy, and as already noted in section 2.3, the far IR couplings easily become dominant as soon as the vW interaction is concerned. Hence, the T → 0 approximation may break down.

A beautiful demonstration of a kind of "spontaneous absorption" of a thermal photon has been given in the Lai and Hinds work (1998), when the decay time from Cs (13S) appears increased as due to an excitation channel to Cs (14P) through a thermal absorption. This process, allowed in the free-space at a non-zero temperature, has been shown to be forbidden when the experiment is performed in too a narrow cavity, when its size is smaller than a wavelength cutoff. Alternately, the thermal field in the vicinity of a surface has been shown to exhibit unexpected coherence properties (see Greffet *et al.*, 2002 and references therein), along with a spectral dependence that is governed by the near-field properties of the blackbody emitter.

Although it is clear that the atom-surface vW interaction can be affected when the energy of the involved virtual atomic transitions falls in the range of thermal photons, only few theoretical works (see *e.g.* Barton, 1997; see also Henkel and Wilkens, 1999) had dealt with this problem. Very recently, the theory suitable for SR-type observation has been developed in our group (Gorza, 2004, in preparation, and Hamdi *et al.*, 2004). One major result is that the resonant coupling between atom and surface, that was limited to an atom virtually emitting into a surface mode, can now be extended to the symmetric process in which the (hot) *surface virtually emits* a photon subsequently absorbed by the atom. The efficiency of this resonant process naturally depends on the thermal population factor { $exp\,(\hbar\omega_{ij}/\,k_BT)\,/\,[1 + exp\,(\hbar\omega_{ij}/\,k_BT)\,]$ }. This surface-emitting process exhibits a notable change of the sign, relatively to the atom-emitting process considered in Eq. (8): to the observed repulsion of Cs($6D_{3/2}$) in front of ($c\perp$) sapphire -associated to the $(6D_{3/2}) \rightarrow (7P_{1/2})$ 12.15 μm emission- , should correspond an enhanced attraction for Cs($7P_{1/2}$) when the temperature is high enough, *i.e.* on the order of ∼ 1000 K. In addition to the thermal effect on the resonant coupling, involving thermal population of surface modes, the non resonant contribution, such as the one appearing in Eq. (7) and featuring an integration over the whole spectrum, is replaced by a discrete summing over the Matsubara frequencies $\omega = kT\,/\,2\pi\,\hbar$. Note that in the principle, the introduction of these thermal excitations introduces new characteristic wavelengths, above which the near-field expansion may become invalid. In spite of this, an elementary $z^{-3}$ interaction remains most often valid, yielding a $C_3(|\,i>)$ value simply dependent on the temperature, *i.e.* $C_3(T)$. The changes induced by this non resonant contribution seem to remain marginal, while the resonant contributions (including those only associated to wings of a resonance), are predicted to be dominant.

Experimentally, it does not seem that these $C_3(T)$ variations have ever been investigated at present. Our SR studies on the $8P_{3/2}$ level of Cs are oriented in view of such an investigation. Indeed, the strong absorption couplings $(8P_{3/2}) \rightarrow (7D_{3/2})$ and $(8P_{3/2}) \rightarrow (7D_{5/2})$ occur respectively at 39 μm and 36 μm, *i.e.* an energy corresponding to the practical temperatures used in SR spectroscopy. With windows in materials such as $BaF_2$ or $CaF_2$ (see Failache 1999, and Fig.18) resonant effect in the atom-surface interaction may appear with temperature, with the vW attraction possibly turned into repulsion for a sufficient window thermal excitation. These predictions assume that the surface modes are known phenomenologically, and that temperature variations on these modes can be neglected, at least for a given range of temperature. In the principle, the real problem is more intricate, because the thermal equilibrium of the bulk material depends itself on its equilibrium at the interface with vacuum. However, the narrowness of atomic resonances should selectively filter these broad couplings to vacuum.

### 5.3 *vW anisotropy and surface-induced inelastic transfer in an atomic beam*

#### 5.3.1 <u>*vW interaction and Metastability transfer*</u>

The intrinsic anisotropy of the surface vW interaction, originating in the quadrupole term in $\mathbf{D}_z^2$, can lead to a variety of *diagonal* interaction for the various Zeeman components, as it has been searched for in the experiments described in section 4.5. The anisotropy of the interaction, with its *non diagonal* contribution, is also susceptible to break down well-established selection rules and enables surface-induced $\Delta J = 2$ transitions. This has led to a search for an inelastic transfer between the two metastable atomic states of rare gases (Ducloy, 1998; Boustimi *et al.,* 2001 a), respectively characterized by the quantum numbers $J = 0$ and $J = 2$.

Due to the large amount of energy involved in such an inelastic process, governed by the same $z^{-3}$ scaling factor, the metastability transfer occurs only at extremely small distances from the wall (that may fall below the "long-range approximation", see section 1). Owing to the long lifetime of metastable atoms, the experimental principle can rely on the mechanical deflection of an atomic beam, enabling one to compare the free atom before the vW interaction, and the output channels for the atoms having undergone such a surface interaction (see Fig. 19). A key point in the detection method is that the change in the internal energy transfer is totally converted into kinetic energy, *i.e.* there is no energy transfer to the surface; moreover, the vW interaction being invariant along the surface, the atom acquires an impulsion transfer **Δp** that is oriented along the surface normal.

**5.3.2** *Experimental observations*

For a monokinetic incident beam of metastable atoms, the extreme accuracy of the energy transfer, combined with the well defined direction for **Δp**, imposes that the atoms having undergone a change of internal state contribute to produce a novel atomic beam, monokinetic and whose angular direction relatively to the incident beam is rigorously known. One has indeed :

$$v_o \cos\theta_o = v_f \cos\theta_f \qquad (19)$$

$$v_f = \left(v_o^2 + 2\Delta E/m\right)^{1/2} \qquad (20)$$

In eqs.(19-20), $v_o$ and $v_f$ are respectively the initial and final atomic velocities, and $\theta_o$ and $\theta_f$ the beam orientations with respect to the surface.

As the energy transfer most often largely exceeds the thermal energy of the atoms (*e.g.* up to 650 meV for Kr, as compared to ~ 70 meV for the velocity-selected thermal energy), the angular deflection is very large (*e.g.* 70°), with the weak dispersion around this value yielding information on the effective state of the surface (*e.g.* lack of planeity as due to local corrugation, breakdown at a microscopic level of the local long-range symmetry,...).

Experimentally, atoms cannot fly at a grazing incidence over a long distance because the interaction is attractive. Rather, the beam is deflected after *transmission* through a thin microslit. Experimental difficulties are many: the counting rates are very low, as due to the low density of metastable states; the impact parameter enabling an atom to undergo deflection is very limited : the typical distance where the metastability transfer is efficient is found to be ~ 5 nm. The shape of the microslit must be well controlled for the interaction to occur during the ~ 100 nm long region where the atom is at grazing incidence. Anyhow, the strong angular selectivity of the process has permitted to observe this signature of vW anisotropy in various situations. The successful observation for metastable Ar and Kr atoms transmitted through a micro-slit (Boustimi *et al.*, 2001 a) was followed by a more efficient process associated to transmission through a nano-grating (Boustimi *et al.*, 2001 b), by an extension to a molecular specie (metastable N$_2$) (Boustimi *et al.*, 2001 c), and even by the observation of the reverse endothermic metastability transfer (Karam *et al.*, 2002).

### 5.4 *Atom interaction with nanobodies*

With the trend to explore shorter and shorter atom-to-surface distances -see sections 5.1 and 5.3-, it becomes notably interesting to consider the interaction of atom with surface featuring "strong" curvatures, *i.e.* a curvature radius smaller than the relevant wavelengths for the atom fluctuations.

Several modifications have to be considered for the interaction of atoms with the nonplanar surfaces (see *e.g.* the review by Klimov *et al.*, 2001):

- The geometrical factor that controls the strength of the electrostatic image from the bulk permittivity ε departs from the common factor [(ε - 1) / (ε + 1)]. It is described in general by a more complex function, that can occasionally be turned into the well-known form factor [(ε - 1) / (ε + 2)] for an atom located in the vicinity of a sphere (Klimov *et al.*, 1996; 1997a, b;1999 a, b) . In particular, the cylindrical limits of ellipsoids have been considered in this problem, with a study of the asymptotic behavior of the corresponding geometrical factors (Klimov and Ducloy, 2000)

- This change in the surface response implies a  frequency shift in the resonant response of the nanobody -relatively to the planar surface response- , when the dielectric medium is dispersive.

- The  z$^{-3}$ interaction law remains valid as long as the atom-surface distance is small compared to the curvature of the nanobody. More generally, the cylindrical symmetry of the interaction can be lost, and extra-anisotropy term naturally appears from the particular geometry of the interacting atom with a specially-shaped nanobody (Klimov *et al.*, 2002, a,b).

- When the curvature radius becomes shorter than the propagation wavelength, the atom-surface interaction is no more limited to a dipole expansion, coupling the atom dipole fluctuations to its induced image. Rather, the complete multipole expansion should be considered. This permits the quadrupole contribution to bring specific resonant contributions, that could turned to be dominant in specific cases (Klimov and Ducloy, 2000).

- A larger variety of situations than the elementary discrimination between attractive and repulsive behaviors can even be considered. In particular, there could exist a possibility  for an atom to orbit around a nanobody (Klimov *et al.*, 1999 b). Specific effects associated with the periodicity of a photonic device are also under investigation.

Until now, the corresponding theoretical developments have not been followed by demonstration experiments, notably because of the lack of a general experimental method for these problems with nanobodies, that would be the equivalent of SR spectroscopy for the interaction of an atom with a plane surface. Note that the experimental developments, in progress in Paris (Treussart *et al.*, 1994; von Klitzing *et al.*, 2001) and in Cal'tech (Vernooy *et al.*, 1998) have until now been more concerned with microbodies (notably microspheres with their specific resonances) whose size largely exceeds the wavelength, than with genuine nanobodies. For a single nanobody, the surrounding volume is very small, and the experimental extension to a distribution of "identical" nanobodies requires homogeneity conditions that are particularly tough to realize with the present state-of-the-art of nanobody production. Also, the atomic motion strongly limits the interaction time. The use of very slow atoms (*i.e.* after laser cooling) could be beneficial, in spite of the already mentioned limitations, and eventually with respect to the quantum nature of the atomic motion and limits due to uncertainty principle at a very small distance from the nanobody. More generally, the resonances of an atom at a very small distance from the surface gets so broad, and the lifetime of its excited level (Klimov *et al.*, 2001) so short, that the advantages of high resolution spectroscopy tends to be limited. Rather, the general formalism developed for such a problem can be applied to an elementary "atomic" system, such as a chromophore, in a more general situation than the one of an atom freely evolving in the space. In particular, these modelings may be of interest for various problems such as the modification of the fluorescence spectrum of an embedded atom (or ion) close to a nanoscope tip (Klimov *et al.*, 2002 b), or a carbone nanotube, or nanofibers (Klimov and Ducloy, 2004)...

Further extensions should deal with the interaction of an atom with a micro- or nano-structured object, such as microstructured fibre, nano-grating, ..., introducing the possibility of tailored shape factor as well as nanobody-driven response, extended over a macroscopic size. Let us also recall that the sensitivity is enhanced when a transmission slit is replaced by a transmission nano-grating in the experiment described in the above section with metastable atoms. It would hence become conceivable that details on the quality of the structured surface are learned from the observed features of the surface interaction.

## 6.    Conclusion

The problem of the long-range electromagnetic coupling, between an atom and a neighboring surface exhibits many facets that the optical methods are well suited to tackle, notably because they provide a natural technique to explore the variety of situations offered by excited atoms. Although the principle description of the vW surface interaction is apparently well-known, the effective experimental exploration has remained limited, with respect to the extremely broad range of energy covered by the $z^{-3}$ interaction. The development of new techniques, enabling complementary exploration of various distance ranges, is hence an important purpose of a fundamental interest, that is partly connected with contemporary quantitative researches on the Casimir effect. Our studies have notably emphasized the importance of the spectral response of the surface, and are potentially sensitive to the various corrections such as anisotropy, lifetime and trajectories modifications, temperature corrections, and surface roughness, that invalidate too elementary predictions. In the same spirit, it should be noted that although the related problem of the vW interaction between two solids is of utmost importance for various problems of biology, like membrane problems, the various estimates derived from simple models have until now revealed too crude to describe sensitively the experimental biological values (Boström *et al.*, 2001). Similarly, one may expect that the understanding of the resonant energy transfer mechanism between molecules (see *e.g.*: Cohen and Mukamel, 2003; Selvin, 2000) could benefit in a fundamental manner from the various observations performed with an atom and a surface, or with gaseous atoms in a confined environment.

**Acknowledgements**

This report has benefited from continuous discussions,  and long-standing co-operation with J. Baudon, M. Fichet, M-P. Gorza, V. Klimov, J.R.R. Leite, V. Letokhov, G. Nienhuis, S. Saltiel, F. Schuller.  The results presented here would not have been obtained over the years without the essential contributions of numerous students, visitors, and co-workers from the staff, and notably A. Amy-Klein, M. Boustimi, M. Chevrollier, G. Dutier, H. Failache, O. Gorceix, M. Gorris-Neveux, I. Hamdi, M. Oriá, N. Papageorgiou, D. Sarkisyan, V. Sautenkov, P.C.S. Segundo, P. Simoneau, T. Vartanyan, A. Yarovitski, A. Weis.

This work contributes to the objectives of the European consortium FASTNet (contract HPRN-CT-2002-00304)

**Figure Captions**

Figure 1 : The electrostatic image induced in a reflector by a  charged dipole system.

Figure 2 : The various ranges of atom-surface distance, with the corresponding type of interaction.

Figure 3: The Cs levels, with the virtual couplings for Cs($6D_{3/2}$) - in black lines - and Cs($6D_{5/2}$) - in grey -. The strength of the couplings, as estimated for an ideal reflector,  -see Table 1- is made visible by the variable thickness of the coupling arrows.

Figure 4 :  (a) $\Re\mathrm{e}[(\varepsilon-1)/(\varepsilon+1)]$  (solid line) and $\Im\mathrm{m}[(\varepsilon-1)/(\varepsilon+1)]$ (dashed line) for sapphire with  the *c*-axis perpendicular to the window (*i.e.* $c_{\perp}$) ; (b) same for fused silica; (c) same for YAG. The data for $\varepsilon$ is extracted from Shubert *et al.* (2000) for sapphire, from Palik (1985) for fused silica, and from Gledhill *et al.* (1991) for YAG.

Figure 5 : Emission in the "forbidden region": when an atomic emitter of the vapor is close enough to the interface, emission in the near-field yields observable fluorescence outside the fluorescence cone defined by  $sin(\theta_{cr}) = 1/n$.

Figure 6 : Tunability of the dielectric image coefficient r (see Eq.(8)) with the orientation of the *c*-axis. The figure illustrates the situation for a sapphire window interacting with an atom whose virtual emission, at 12.15 µm, is in resonance with the surface modes. θ is the angle between the *c*-axis and the normal to the window. The atomic state itself is assumed to be isotropic.

Figure 7 : The principle of Selective Reflection (SR) spectroscopy is the detection of a resonant change  (*i.e.* when the irradiating frequency $\omega$ is close to the vapor resonance at $\omega_0$) $\Delta R(\omega)$ in the field reflected at the window interface. This change originates in the field radiated, for each "slice" of the medium, by the oscillating dipole $\mathbf{p}(z)$ induced by the irradiating field.

Figure 8 : The theoretical SR lineshape, in the local model (dispersive Voigt profile, in dashed line), and in the nonlocal model (solid line). The horizontal frequency scale is in Doppler width (ku) units, the homogeneous width  is 0.01 times the Doppler width.

Figure 9 : Theoretical frequecy-modulated (FM) SR lineshapes taking into account the strength of the vW interaction through the dimensionless A parameter (see Eq. 15). The calculation is "universal", as due to the infinite Doppler width approximation.

Figure 10 : The principle set-up for evanescent wave spectroscopy.

Figure 11 :  Spectroscopy in a thin cell: for a given modulus of the velocity (*e.g.* most probable thermal velocity), longer atomic trajectories and a longer interaction time with the laser light are allowed for atoms with a small velocity along the normal to the windows (*i.e.* small $v_\perp$ ). Note that the laser beam diameter is assumed to exceed largely the cell thickness L.

Figure12 : A typical direct SR spectrum (*i.e.* without FM), recorded on the Cs $D_2$ resonance line $6S_{1/2}(F = 4) \rightarrow 6P_{3/2}(F'=3,4,5)$.

Figure 13 : An example of the fitting (narrow black line) of a FM SR lineshape (in grey). The fitting includes here 3 different hyperfine components, and is performed with a single parameter for the vW strength parameter (and a single homogenous optical width). It also takes into account the frequencies of the unshifted (free-space) resonance, as revealed from the Saturated Absorption (SA) reference. The present FM SR lineshape is obtained in the situation of a repulsive interaction between Cs ($6D_{3/2}$) and a sapphire ($c_\perp$) window (see Failache *et al.*, 1999 and 2003).

Figure 14 : The consistency of the vW fitting over a pressure-induced modification of the resonance width. One has simultaneously extrapolated from the fittings a simultaneous linear pressure broadening, and a vW strength independent of he pressure. The FM SR experiments are performed in the situation of a repulsive interaction between Cs ($6D_{3/2}$) and a sapphire ($c_\perp$) window, with spectra such as the one shown in Fig 13.

Figure 15 : Scheme of an experiment aiming at detecting anisotropy effects in vW shift, with a resolved Zeeman structure. The two lasers have a similarly polarization, perpendicular to the magnetic field **B**, so that they induce inside the vapor (bulk) an identical interaction. In SR spectroscopy, there remains a cylindrical symmetry for the vW interaction for the experiment with the (//)laser (**B** along the surface normal), that is destroyed for the ($\perp$) laser (**B** parallel to the surface.

Figure 16 : FM reflection on the Cs $D_1$ line in a 80 nm thick ETC. The dashed lines indicate the position of the volume resonances -as obtained from a SA reference. The  large observed

shift between ETC resonances, and volume resonances, illustrates the signature of the vW interaction.

Figure 17 : The energy level scheme of Cs that could lead to a surface-induced level-crossing in the vicinity of a sapphire window ($c_\perp$). The $6D_{3/2}$ level is strongly repelled by the sapphire surface, while the $6D_{5/2}$ level undergoes an ordinary vW attraction.

Figure 18 : $\Re e\left[(\varepsilon-1)/(\varepsilon+1)\right]$ for a $BaF_2$ window (solid line), and a $CaF_2$ window (dashed line). The predicted surface resonances should enable resonant coupling of a (virtual) atom absorption with a thermally excited surface. The data (see Failache (1999)) has been extrapolated from the bulk values as given in Kaiser *et al.* (1962).

Figure 19 : The principle of the experiment for the detection of a surface-induced metastability transfer. The detector of metastable states (*e.g.* initial state $^3P_0$ , or $^3P_2$ transfer state) can be rotated from the direction of the incident atomic beam, to the direction $\theta_f$. The incident beam is here oriented parallel to the interacting surface, *i.e.* $\theta = 0$. The region of surface interaction extends over a length up to 50 $\mu$m (for a slit) down to 50 nm (for a nano-grating) (*cf.* Boustimi *et al.*, 2001).

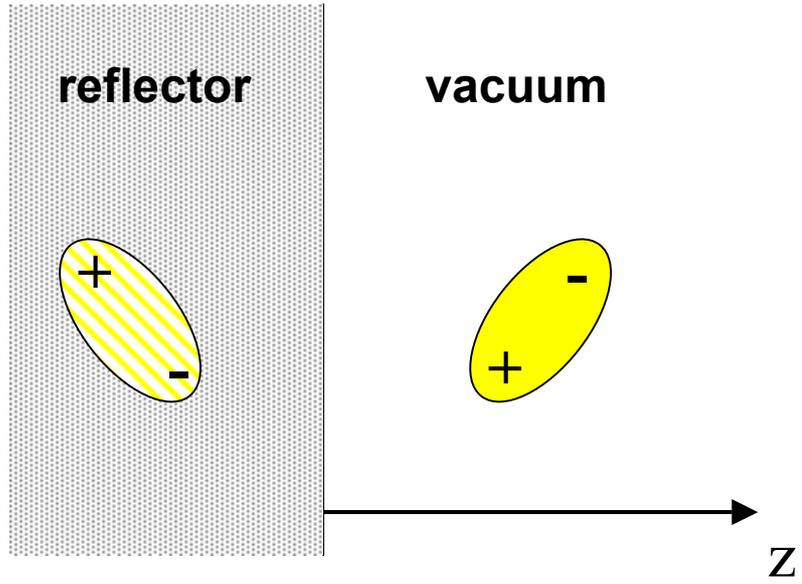

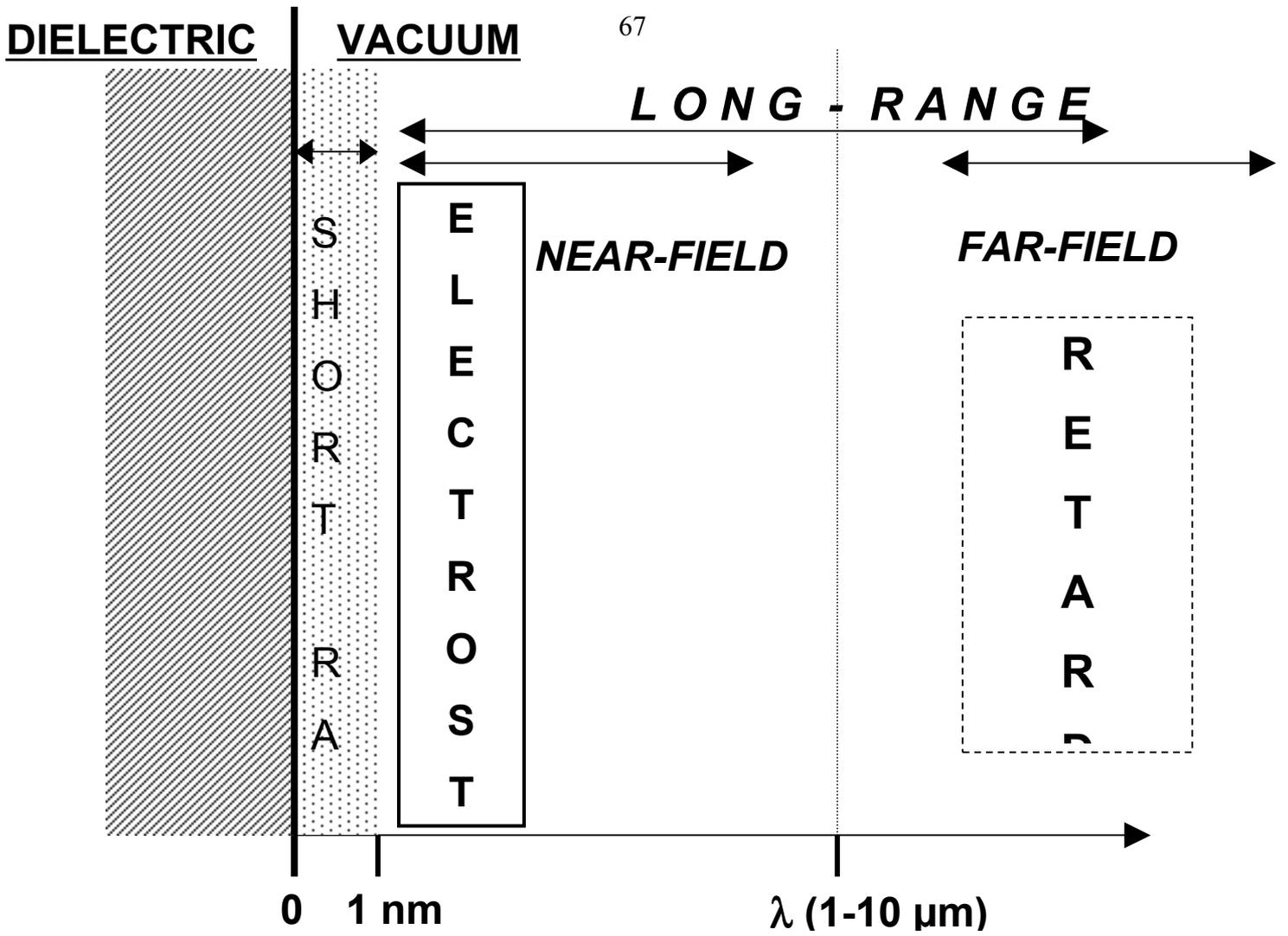

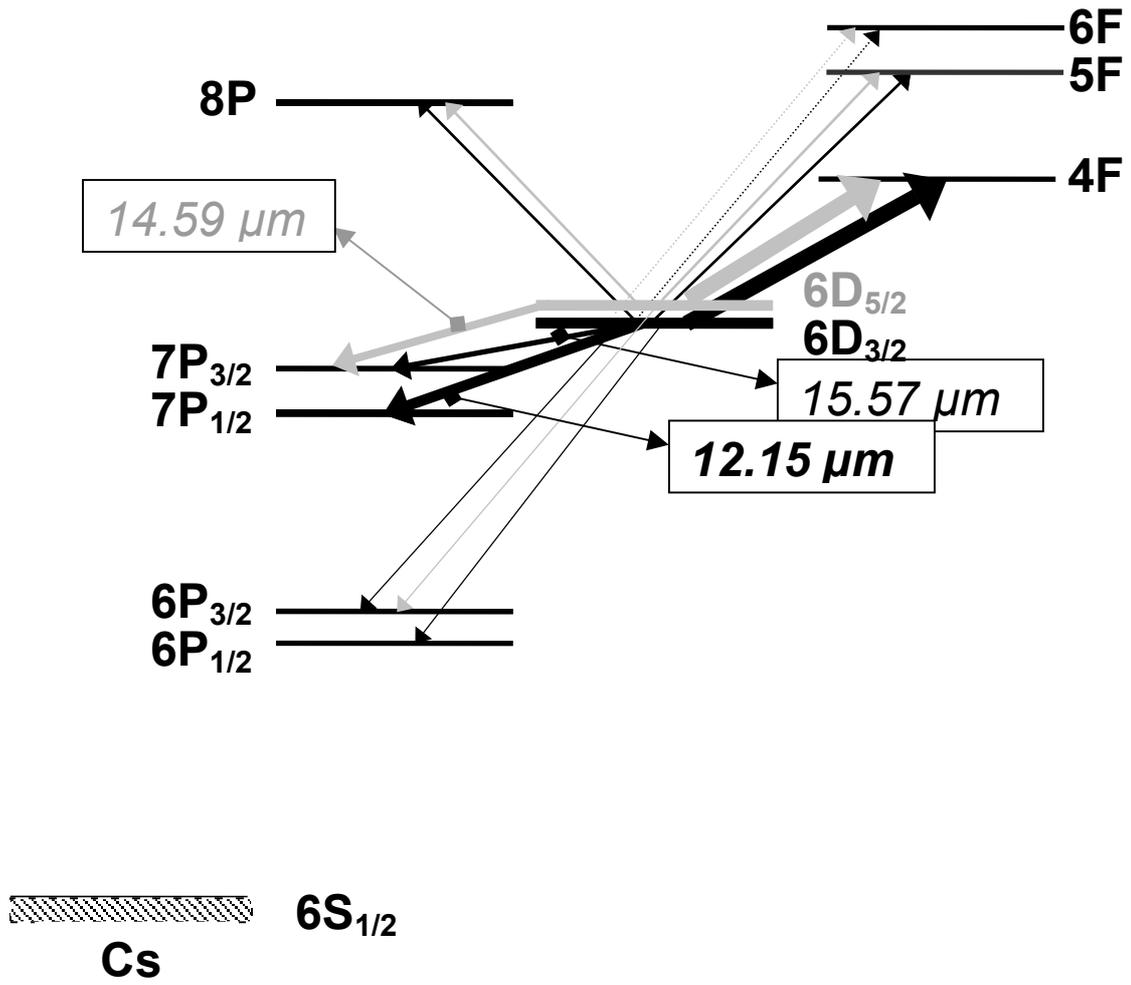

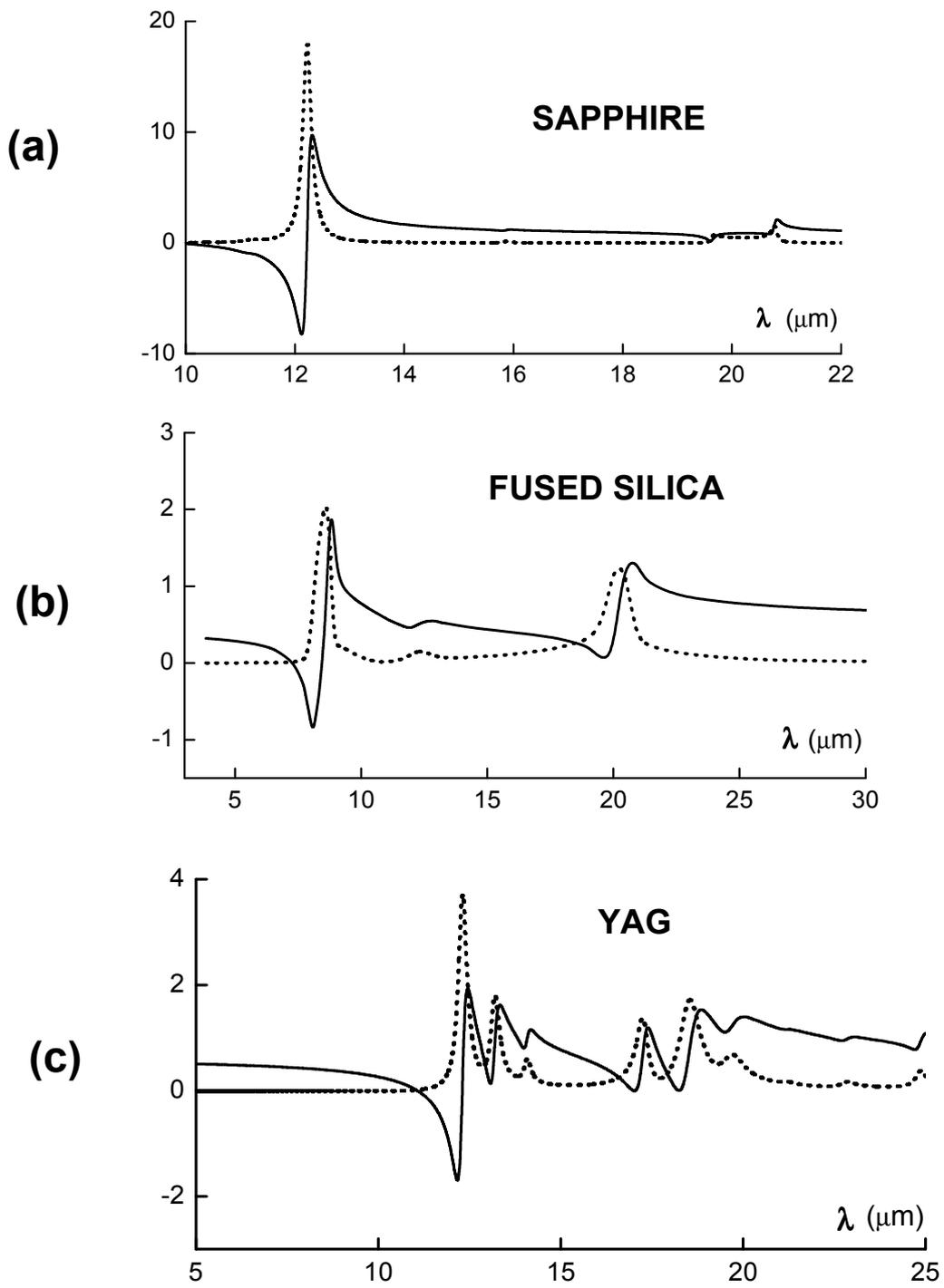

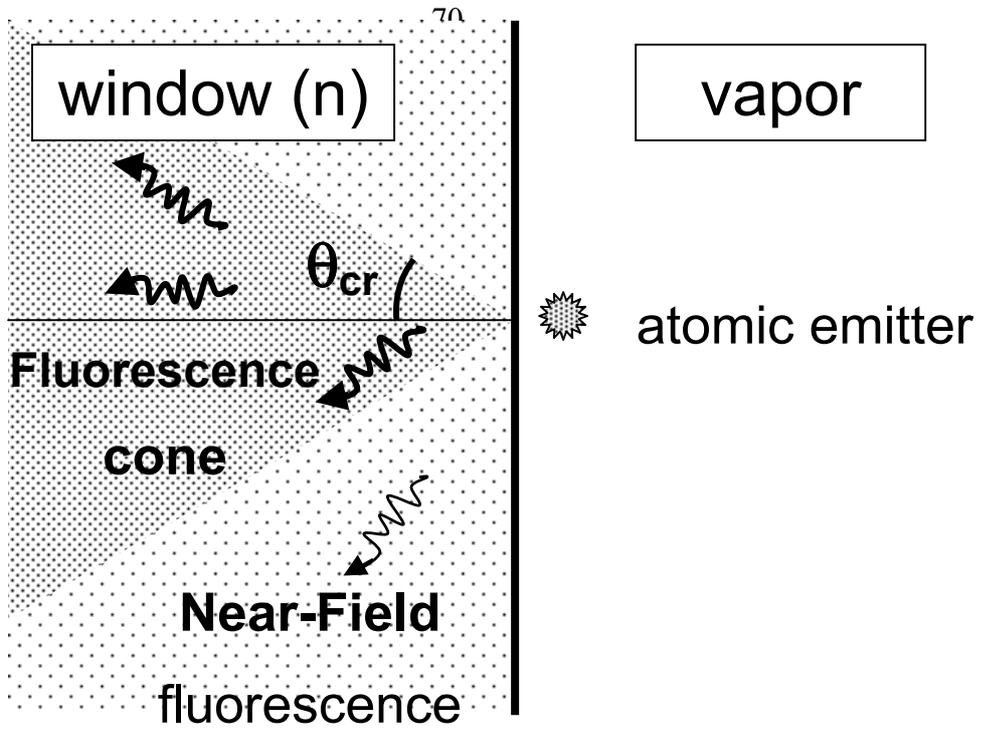

window (n)

vapor

$\theta_0$

$\theta_{cr}$

atomic emitter

**Fluorescence**

**cone**

**Near-Field**

fluorescence

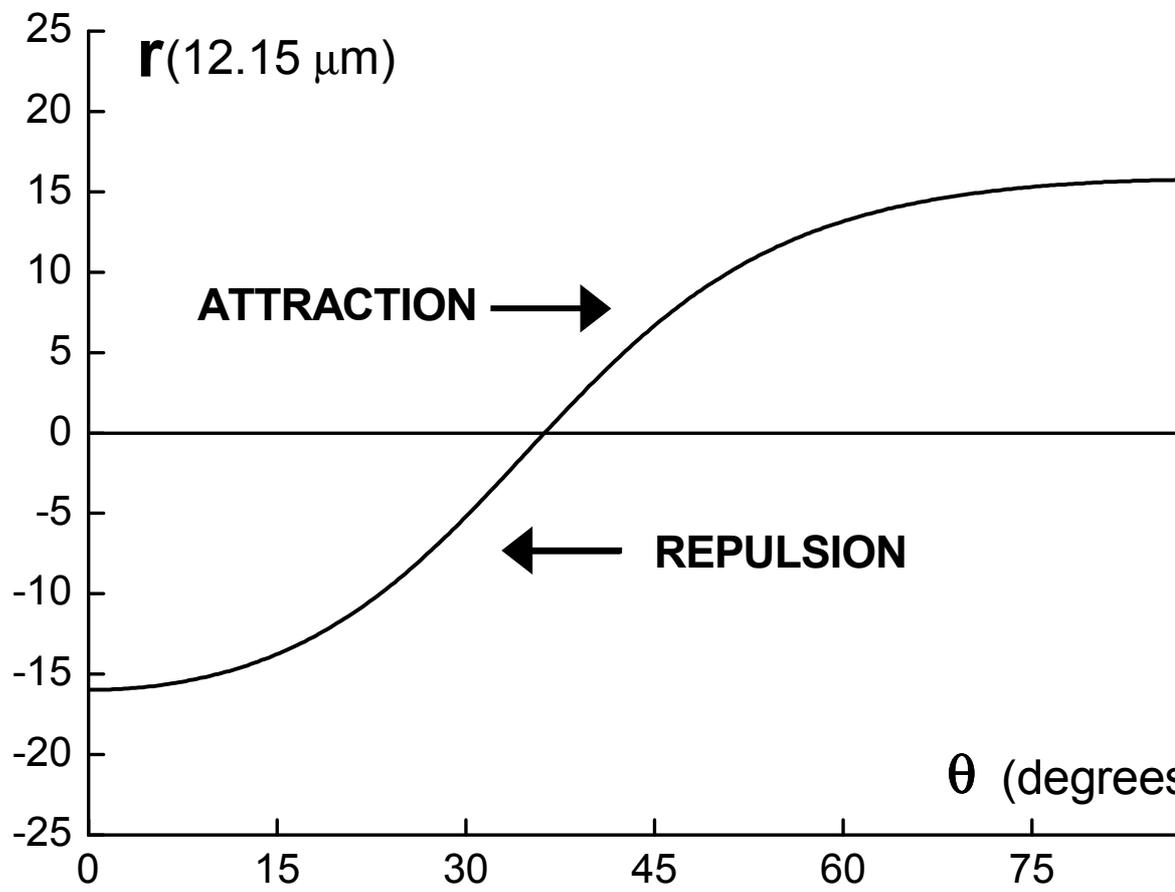

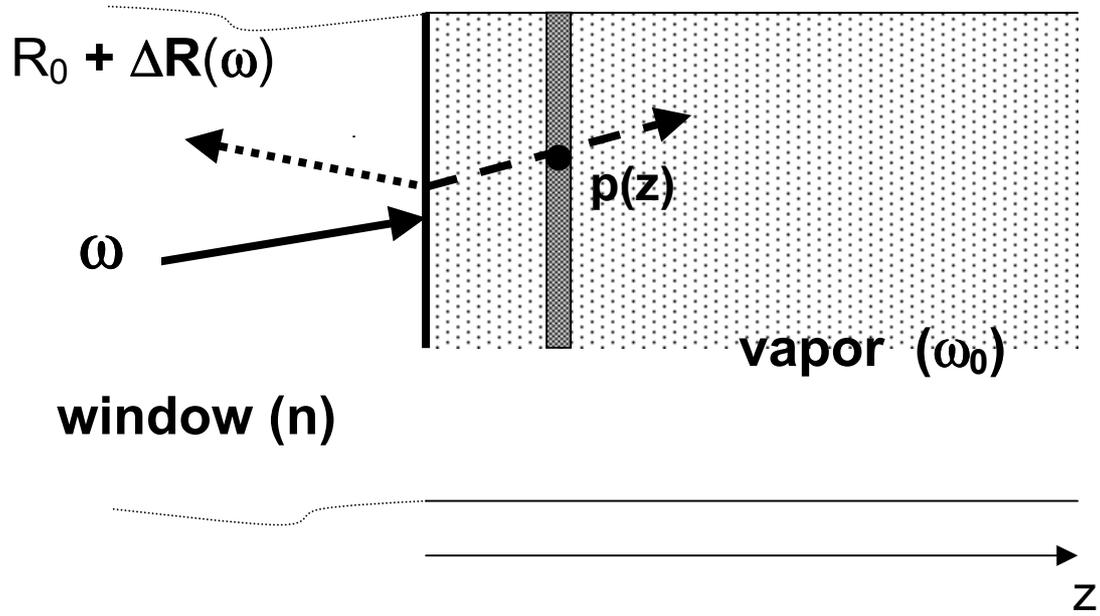

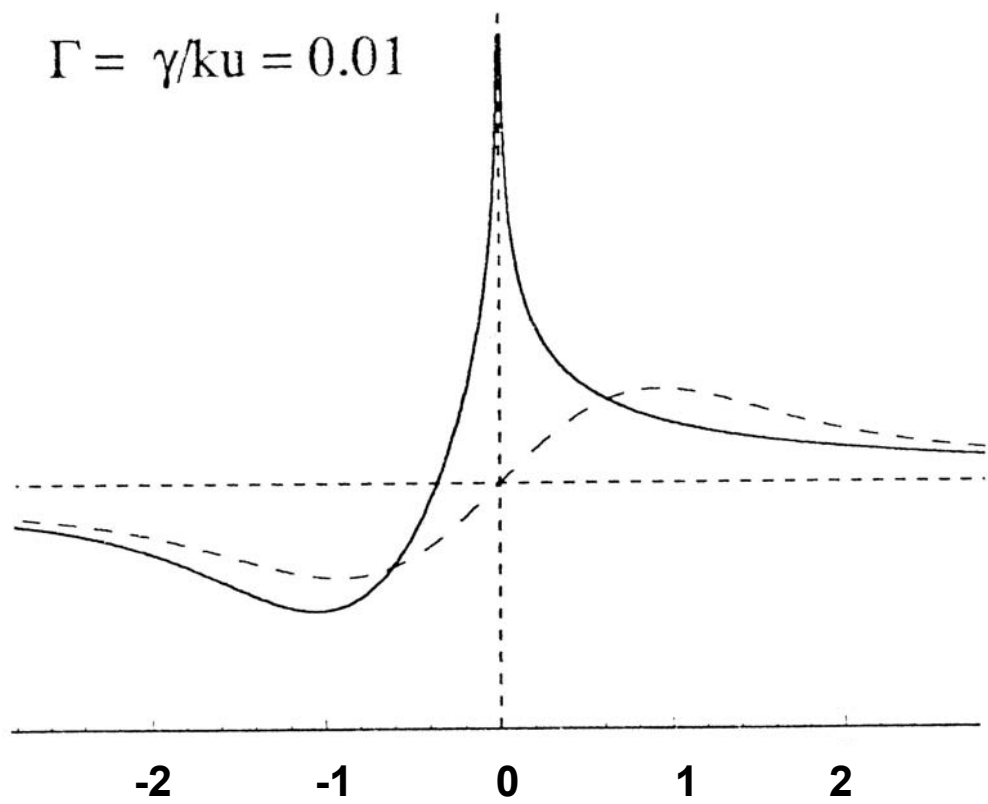

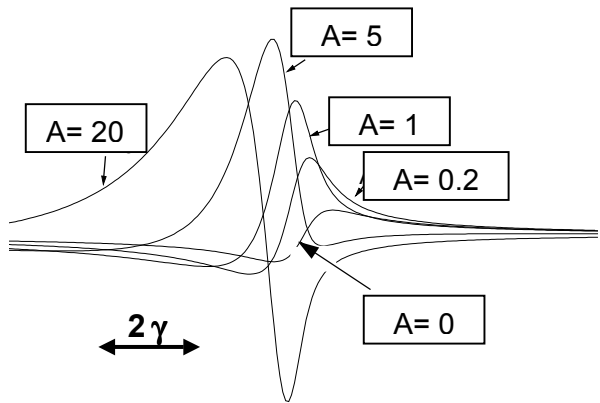

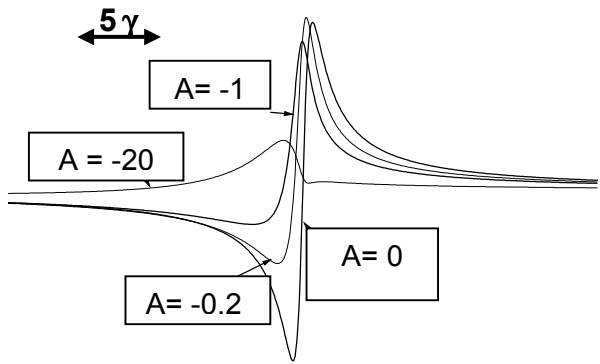

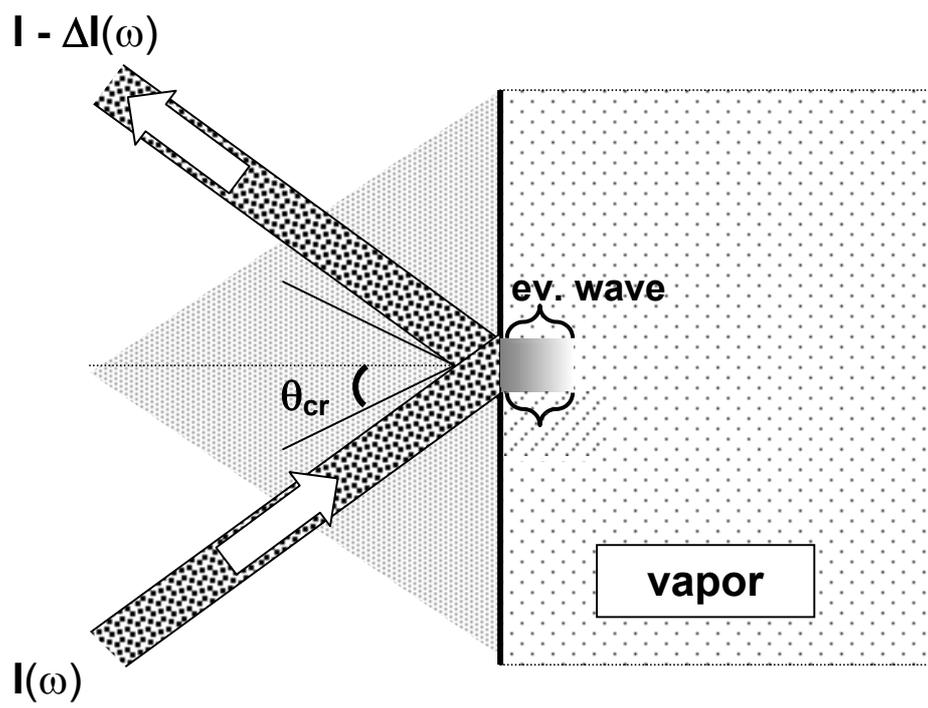

I - ΔI(ω)

ev. wave

$\theta_{cr}$

vapor

I(ω)

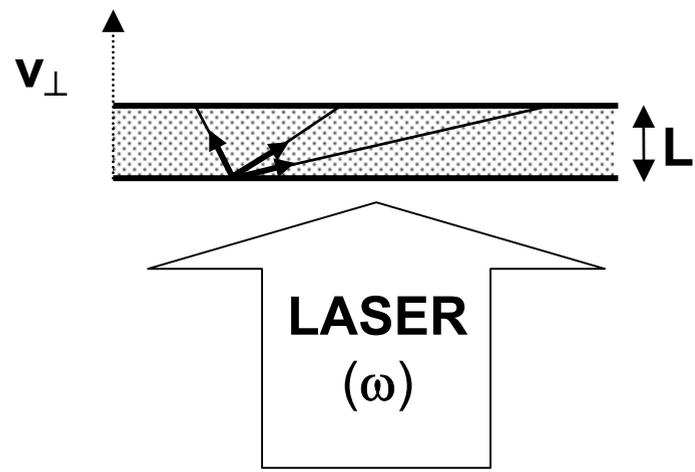

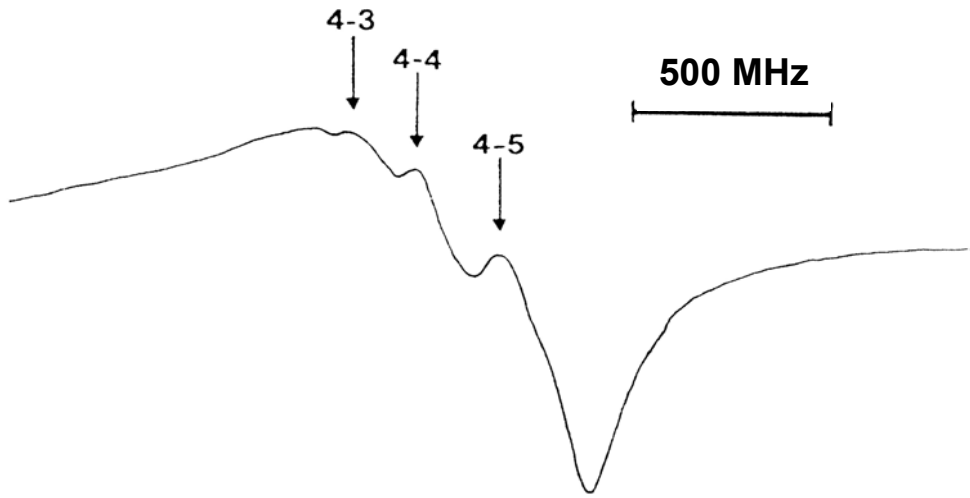

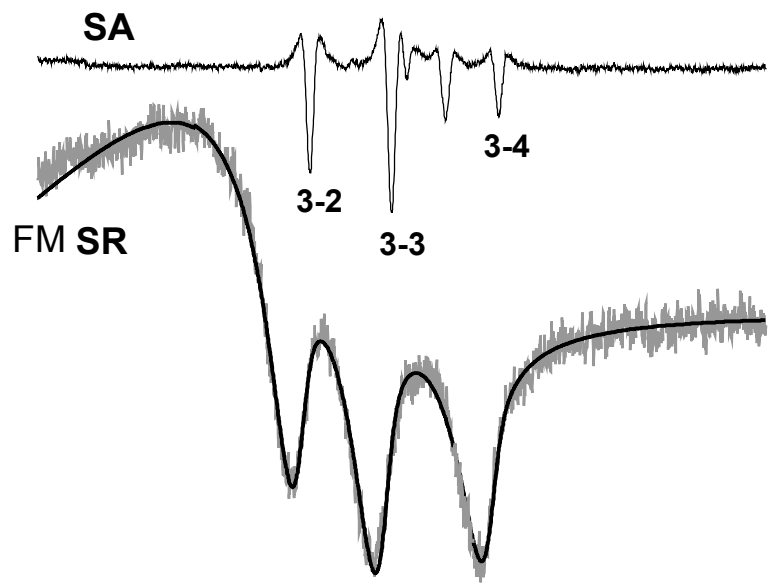

**SA**

**3-2**

**3-3**

**3-4**

FM **SR**

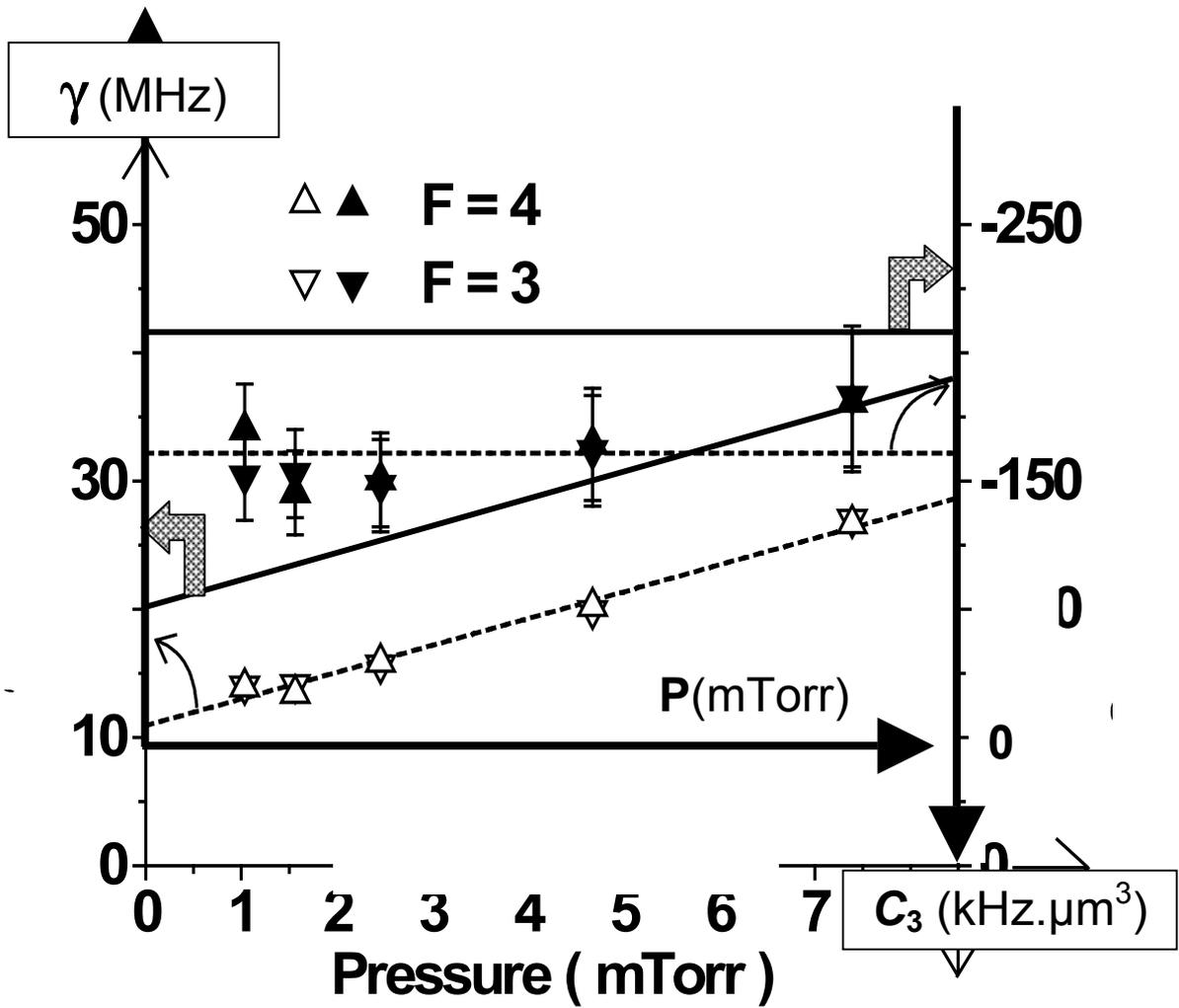

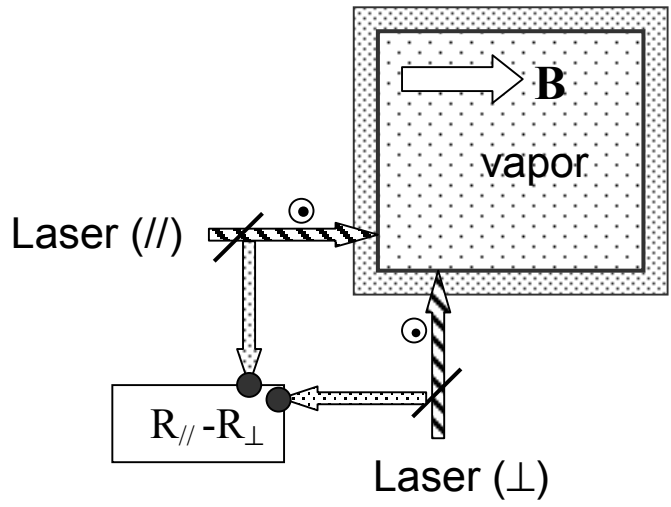

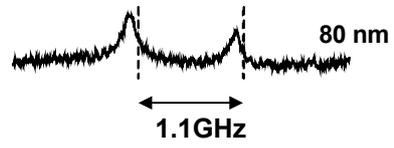

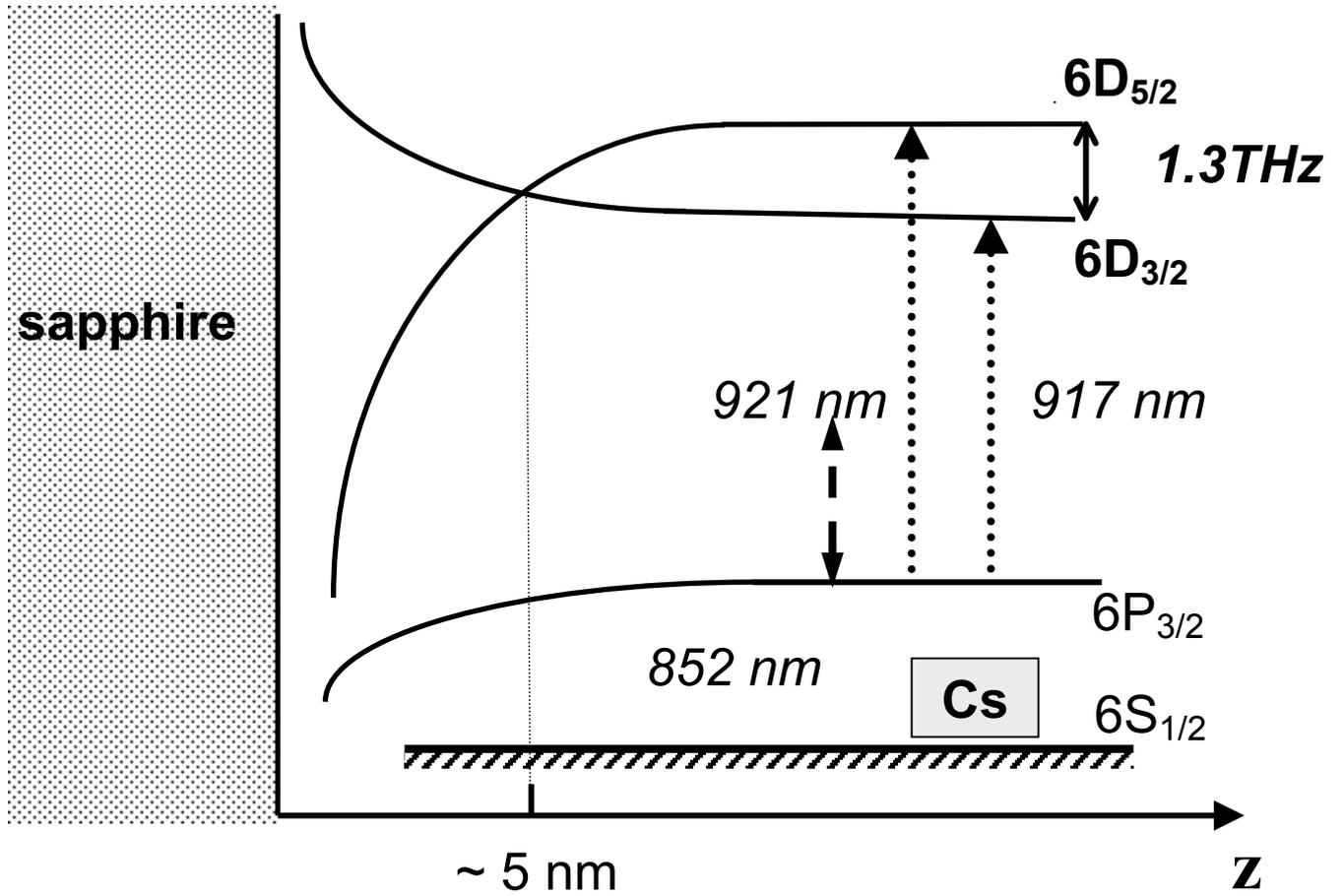

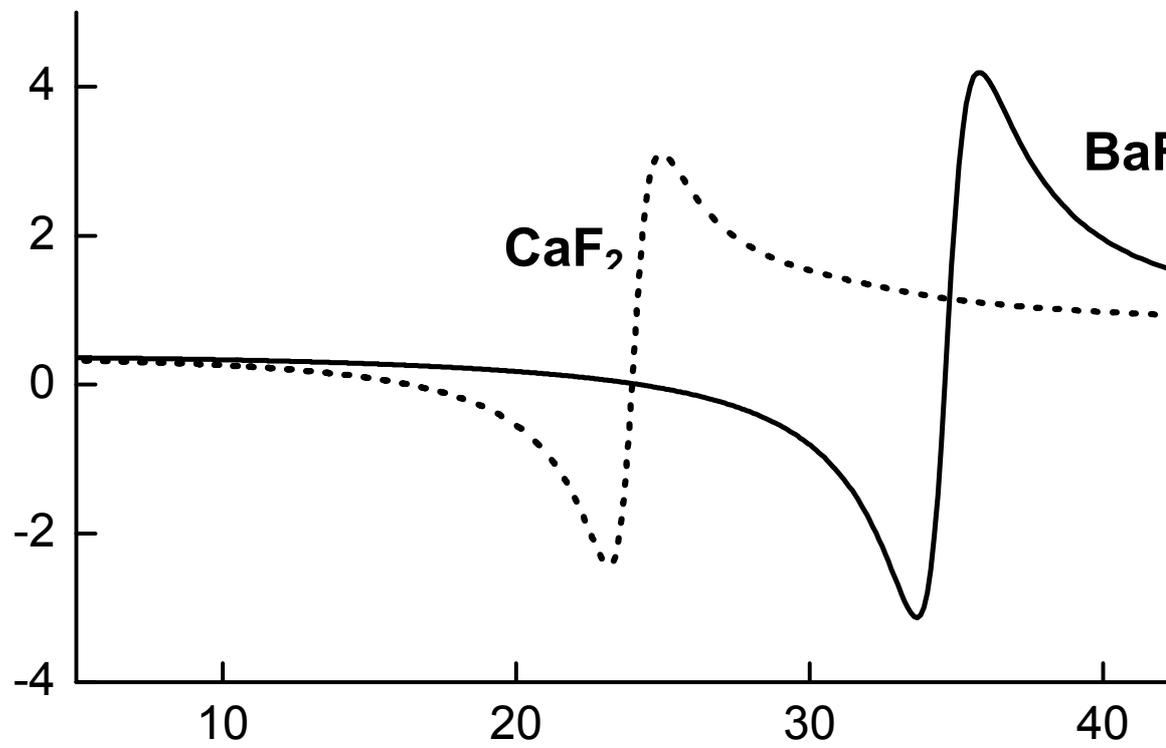

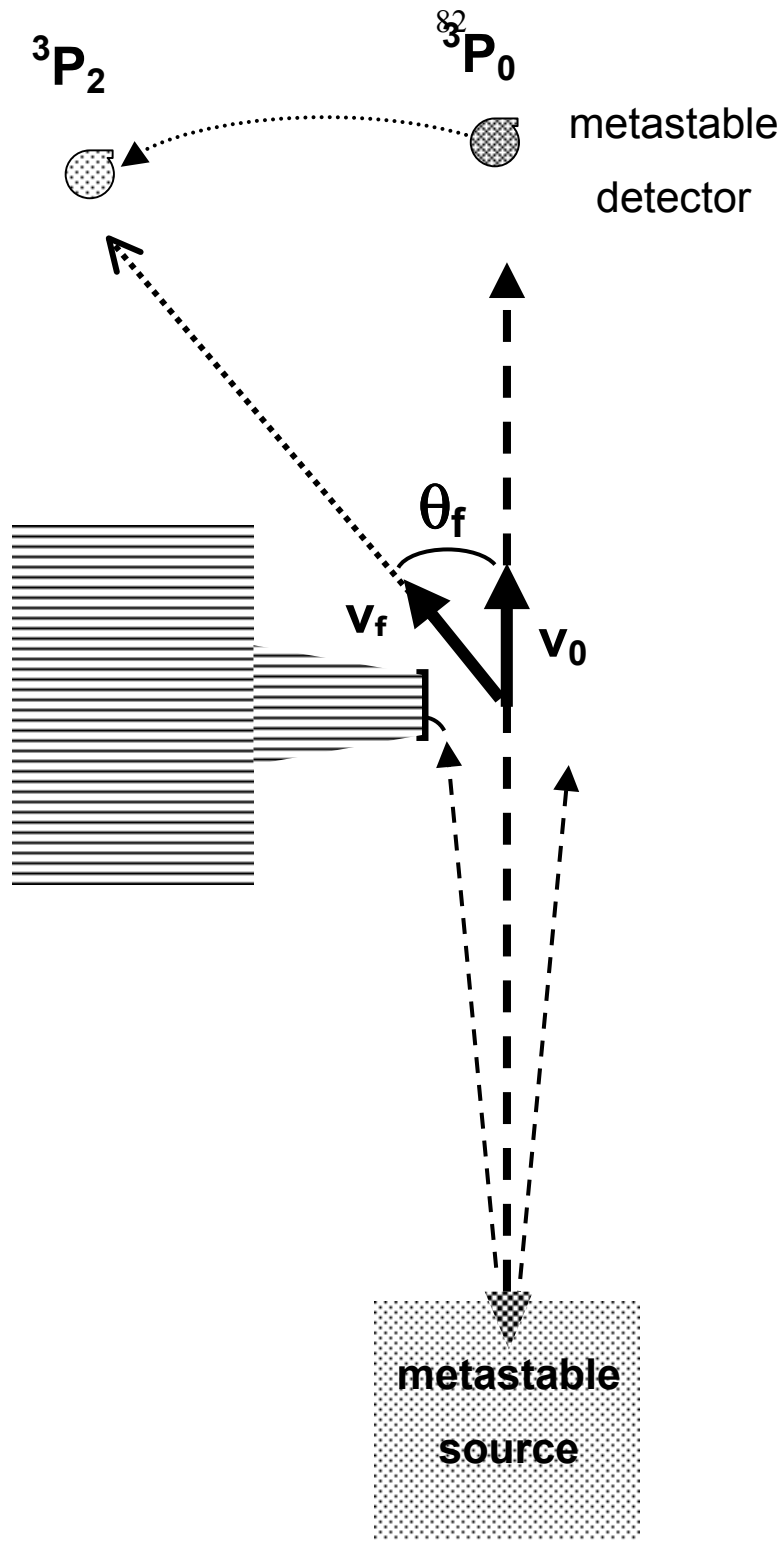

$^3P_2$       $^{88}_{38}P_0$    metastable

detector

$\theta_f$

$v_f$    $v_0$

**metastable**

**source**